\documentclass[aps,pra,reprint,superscriptaddress,floatfix,showpacs,showkeys]{revtex4-2}
\usepackage{xcolor}
\usepackage{graphicx}
\usepackage{dcolumn}
\usepackage{siunitx}
\usepackage{float}
\usepackage{bm}

\begin{document}
\title{Engineering Purcell factor anisotropy for dark and bright excitons in two dimensional semiconductors}

\author{Lekshmi Eswaramoorthy}
 \affiliation{Laboratory of Optics of Quantum Materials, Department of Physics, Indian Institute of Technology Bombay, Powai, Mumbai -- 400076, India}
 \affiliation{Department of Materials Science and Engineering, Monash University, Clayton, Victoria 3800, Australia}
 \affiliation{IITB-Monash Research Academy, Indian Institute of Technology Bombay, Powai, Mumbai -- 400076, India}

\author{Sudha Mokkapati}
 \affiliation{Department of Materials Science and Engineering, Monash University, Clayton, Victoria 3800, Australia}
 
\author{Anshuman Kumar}
\email{anshuman.kumar@iitb.ac.in}
 \affiliation{Laboratory of Optics of Quantum Materials, Department of Physics, Indian Institute of Technology Bombay, Powai, Mumbai -- 400076, India}%
\date{\today}

\begin{abstract}
  Tightly bound dark excitons in atomically thin semiconductors can be used for various optoelectronic applications including light storage and quantum communication. Their optical accessibility is however limited due to their out-of-plane transition dipole moment. We thus propose to strengthen the coupling of dark excitons in two dimensional materials with out-of-plane resonant modes of a cavity at room temperature, by engineering the anisotropy in the Purcell factor. A silica micro-disk characterised by high confinement of light in small modal volume, high Q-factor and free spectral range is used to couple to the excitons in monolayer transition metal dichalcogenides. We show numerically that the tapering of sidewalls of the micro-disk is an extremely versatile route for achieving the selective coupling of whispering gallery modes to light emitted from out-of-plane dipoles to the detriment of that from in-plane ones for four representative monolayer transition metal dichalcogenides.     

\end{abstract}

\maketitle

\section{\label{sec:level1}Introduction}

Two dimensional (2D) semiconductors are of great interest for photonic and optoelectronic applications due to their unique optical and electronic properties \cite{Khan2020a,Glavin2020a}. Within this class, the 2D transition metal dichalcogenide (TMDC) materials exhibit direct bandgaps over a wide range from visible to near-infrared with strong excitonic emission \cite{Rasnok2014NanophotonicsDichalcogenides}. The strong confinement of the charges in 2D TMDCs results in high binding energy ($E_b$) of the excitons which is about 1-2 orders of magnitude greater than that in quasi-2D systems such as GaN, GaAs quantum wells\cite{Mueller2018ExcitonSemiconductors}. Such large values of binding energies of excitons make them dominate the optical properties of 2D TMDCs and allow the observation of various classes of excitons including the neutral, charged, localized, dark, intervalley, excitonic molecules and many other bound states\cite{Yuan2017ExcitonSemiconductors}. The optical characteristics of these monolayer TMDCs are widely tunable by varying various parameters like doping, thickness and the dielectric and photonic environment making them excellent choices for optoelectronic devices\cite{Mueller2018ExcitonSemiconductors}. 

For a photon with energy $\hbar \omega$ and $\mathbf{q}_{\|}$ as the projection of momentum on the TMDC plane, the optical selection rules allow the transitions which are conserved in energy ($\hbar \omega=E_{2}-E_{1}$ where $E_{1}$, $E_{2}$ are energy states), momentum ($\mathbf{p}_{\|}=\mathbf{p}_{h}+\mathbf{p}_{e}$, where $\mathbf{p}_{h}$ is the momentum of a hole in the valence band and $\mathbf{p}_{e}$ is the momentum of an electron in conduction band) and spin (between states with the same spin projection). These transitions result in the formation of the bright ($X_0$) or the optically active excitons, while the forbidden transitions result in dark excitons ($X_D$)\cite{Oody2016Exciton,Jiang2021Real-timeDichalcogenide,Selig2018AndDichalcogenides}. Though forbidden, they can be excited by their interactions with bright excitons, an external electric or magnetic field\cite{LohBrighteningArrays}. These dark excitons have relatively long radiative lifetimes compared their bright counterparts due to suppressed direct photon emission, making them ideal candidates for light  storage and sensing applications \cite{Meng2020ProbingExciton}. But, their out-of-plane dipole moment makes their optical accessibility limited. The brightening of these dark excitons has been explored via numerous methods. One such being magnetic brightening where a magnetic field of large magnitude (usually above 10T) is applied which alters the spin alignment without drastically disturbing the electronic structure \cite{Lu2019,PhysRevLett.123.096803,Zhang2017,Molas2017,PhysRevB.96.155423,RobertMeasurementExcitons}. An application of an out-of-plane electric field via plasmon polaritons can probe the dark excitons through strong near-field coupling \cite{Zhou2017,Park2017}. Photoluminescence collection from the monolayer sample edge\cite{PhysRevLett.119.047401} and angle resolved measurements\cite{Schneider2020} were also employed to detect the in-plane emission of dark excitons.

\begin{figure*}[ht]
    \centering
    \includegraphics[width=1\linewidth]{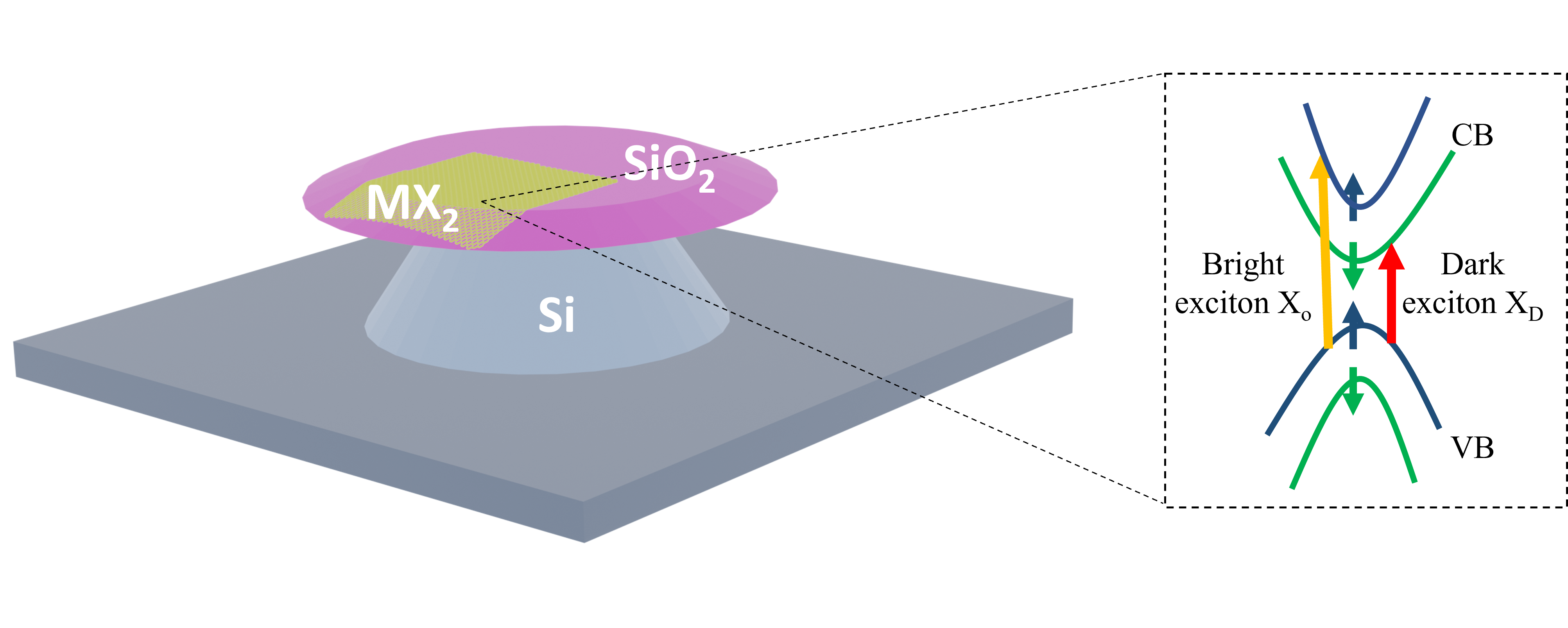}
    \caption{Schematic of proposed methodology of selectively coupling excitons of 2D TMDCs to the planar silica microdisk with (inset) the excitonic transitions in 2D TMDC depicting the bright and spin-forbidden dark excitons}
    \label{fig:1}
\end{figure*}

An alternative way to probe these dark excitons is to alter the exciton dynamics using engineered microcavities. The exciton dynamics can be best explained using a two-level quantum model where an excitonic state is described by an excitation frequency and a radiative decay rate. Further refinements include decoherence rate, intervalley scattering and other non-radiative decay channels. It is well known that by engineering the photonic environment of such a quantum emitter one can tune the local optical density of states thereby engineering the radiative decay rate{\cite{Purcell1946, PhysRevLett.58.2059}}. For instance, R Khelifa et al. \cite{Khelifa2020CouplingHeterostructures} demonstrate waveguide-coupled disk resonators made of hexagonal boron nitride (h-BN) to selectively couple emission from interlayer excitons of a MoSe2–WSe2 heterobilayer, which also show an out of plane dipole component. The high oscillator strengths of intralayer transitions of the two monolayers lead to a high absorption efficiency. On the other hand, the low oscillator strength of the interlayer transition suppresses the reabsorption at the emission wavelength. The combination of these two effects gives rise to strong interlayer emission from the vdW disk resonators. Tobias Heindel et al. \cite{Heindel2018AccessingMicrolenses} exploit deterministically fabricated quantum-dot microlenses to boost photon extraction and access the dark excitons through spin-blockaded metastable biexcitons. D Andres-Penares et al. \cite{Andres-penaresOut-of-planeMicroresonators} propose a versatile spectroscopic method that enables the identification of the out-of-plane component of dipoles due to the selective coupling of light emitted by in-plane and out-of-plane dipoles to the whispering gallery modes of spherical dielectric microresonators, in close contact to them. The proposed method relies on the different selective abilities of IP and OP dipolar emission to excite whispering gallery modes of silica microspheres deposited on top of the emitting layers. While the approaches mentioned above have enabled access to dark excitons, they do not seem feasible for a practical realisation in the fabrication schemes owing to their requirements of cryogenic temperature or large magnetic fields and complex geometry.

We propose to utilise a silica microdisk resonator platform to selectively couple the dark and bright excitons by engineering the resonator geometry. Whispering gallery modes (WGM) resonators allow light to propagate along their circumference, making evanescent coupling of light in and out of the cavities facile \cite{Liu2021NonlinearFingerprinting, Vahala2005}. These are a class of resonators characterised by high Q factors, low modal volume and high free spectral range (FSR). Silica as a choice of resonator material allows for reduced optical loss and easy chip integration \cite{Ferrera2008Low-powerStructures,Miya2000Silica-BasedAnd}. Silica microdisk has one of the simplest geometries of the WGM resonators and is robust to fabrication imperfections \cite{Ying2018ComparisonPhotonics}. The central idea in our proposal is the engineering of the wedge angle of a tapered microdisk. The tapering of the microdisk is a result of the wet etching process used in the fabrication scheme of the silica microdisk\cite{Li2015High-Chip}. {The resonant modes tend to shift away from the sidewalls of the microcavity with increase in inclination thereby reducing radiation loss. This proves advantageous in increasing the Q-factor and Finesse of the cavity \cite{Chen2018EffectsLasers}.} 
It is also reported that the tapering of microdisks results in compensation of the normal dispersion of the cavity \cite{Iyang2019VisibleAngle}. We model the 2D TMDC excitons as dipoles and study the effect of in-plane ($IP$) and out-of-plane ($OP$) dipole orientation on their selective coupling to the WGMs of the tapered microdisk. Thereby the relative contributions of dark and bright excitons of the 2D TMDCs are analysed. The dark and bright excitons have predominantly out-of-plane and in-plane dipole moments respectively. 

\begin{figure}[hb]
    \centering
    \includegraphics[width=0.95\linewidth]{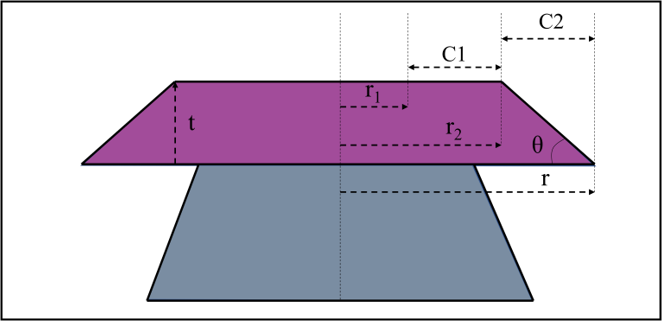}
    \caption{Schematic of the cross-section of microdisk along with the depiction of geometrical parameters radius ($r$), thickness ($t$) and taper angle ($\theta$)}
    \label{fig:2}
\end{figure}

\begin{figure*}[ht]
    \centering
    \includegraphics[width=\linewidth]{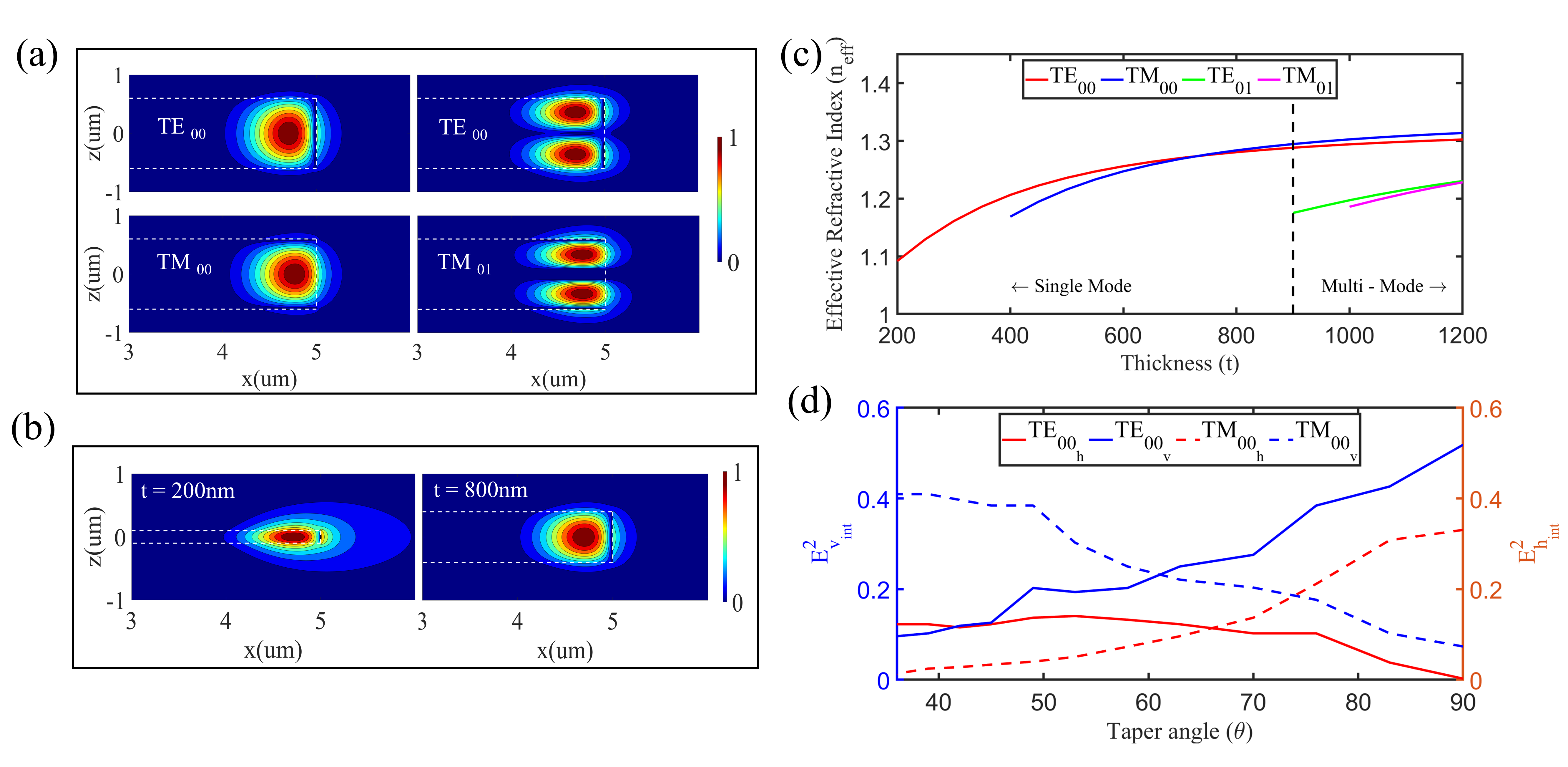}
    \caption{(a) Electric field intensity profiles (normalised to 1) of fundamental $TE_{00}$, $TM_{00}$ and the higher order $TE_{01}$, $TM_{01}$ modes  supported by microdisk of $t = 1200\si{nm}$; (b) $TE_{00}$ mode supported by $t = 200\si{nm}$ and $t = 800\si{nm}$ respectively; (c) Modal analysis of microdisks of varying thickness $t$ showing the dependence of effective refractive index ($n_{eff}$) of the permissible modes on $t$; (d) Dependence of $E^2_{h_{int}}$ and $E^2_{v_{int}}$ over taper angle $\theta$ illustrating Purcell anisotropy at a given $\theta$} 
    \label{fig:3}
\end{figure*}

\section{Results and discussion}

WGM microdisk resonators have significant morphological dependence. Radius ($r$), thickness ($t$) and wedge angle ($\theta$) of the sidewall of a microdisk (Fig.~\ref{fig:2}) are important parameters that alter the propagation of light within the cavity. In the following, we fix the radius of the silica microdisk at $\SI{5}{\um}$, though the analysis can be easily extended to larger disk sizes. The TMDC layer is deposited on top of the microdisk and covers part of the tapered region as shown in Fig.~\ref{fig:1}.

\subsection{Thickness parameter}
The thickness ($t$) of the microdisk is varied between $200\si{nm}$ and $1200\si{nm}$ and the permissible modes corresponding to each of these thicknesses are analysed using the Finite-Difference Eigenmode (FDE) solver. Fig.~\ref{fig:3}(c) shows the variation of the effective refractive index ($n_{eff}$) for these allowed modes of the microdisk of varying $t$. The microdisks of $t < 900\si{nm}$ support only the fundamental $TE_{00}$ and $TM_{00}$ which are shown in Fig.~\ref{fig:3}a thus making the thickness range $200\si{nm} < t < 900\si{nm}$ constitute the single-mode operational region of the microdisk. While the microdisk with $t \geq 900\si{nm}$ supports the higher-order modes such as the $TE_{01}$ and $TM_{01}$ as shown in Fig.~\ref{fig:3}a clearly making it the multi-modal region of the microdisk. The electric field profiles of the fundamental $TE_{00}$ mode illustrate that the mode is clearly evanescent for smaller values of $t$ such as $t = {200}\si{nm}$ as plotted in Fig.~\ref{fig:3}b. While the mode remains bounded within the disk boundary in case of $t = {800}\si{nm}$.
We thus choose $t = 800\si{nm}$ which supports bounded mode with the reduced modal competition. 

\subsection{Tapering for selective coupling}

The dependence of $\theta$ (angle subtended by the sidewall and the bottom side of the microdisk with $\theta = 90^0$ representing a normal sidewall and $\theta < 90^0$ representing a tapered sidewall as shown in Fig.~\ref{fig:2} on the excitation of WGMs of the microdisk is quantified using Purcell factor ($PF$) which is a measure of the interaction strength of excitons with light. The decay rate of an emitter oscillating at a frequency $\omega$ in free space with a dipole moment $d$  is given by $\Gamma_0$\cite{Novotny2006}:
\begin{equation}
\Gamma_{0}=\omega_{0}^{3}\left|\mathbf{d}\right|^{2} /\left(12 \pi \varepsilon_{0} \hbar c^{3}\right)
\label{eq:1}
\end{equation}
The resonant cavity alters the decay rate of the emitter without perturbing the frequency of emission, since the Lamb shift is usually quite small compared to the resonance frequency\cite{Park2016}. Fermi's Golden rule state that $\Gamma$ is proportional to the local density of states (LDOS) which is dependent on the position, orientation and transition frequency of the emitter dipole. $PF$ defines the ratio of decay rate of the emitter in the near field of the cavity ($\Gamma_c$) to that in free space ($\Gamma_0$) as given in Eq.\ref{eq:2}
\begin{equation}
PF = \frac{\Gamma_c}{\Gamma_{0}} = 1+\frac{6 \pi \varepsilon_{0}}{\left|\mathbf{d}\right|^{2}} \frac{1}{k^{3}} {\Im}\left[\mathbf{d}^{*} \cdot \mathbf{E}_{\mathrm{s}}\left(\mathbf{r}_{\mathrm{e}}\right)\right]
\label{eq:2}
\end{equation}
\begin{figure*}[ht]
    \centering
    \includegraphics[width=1\linewidth]{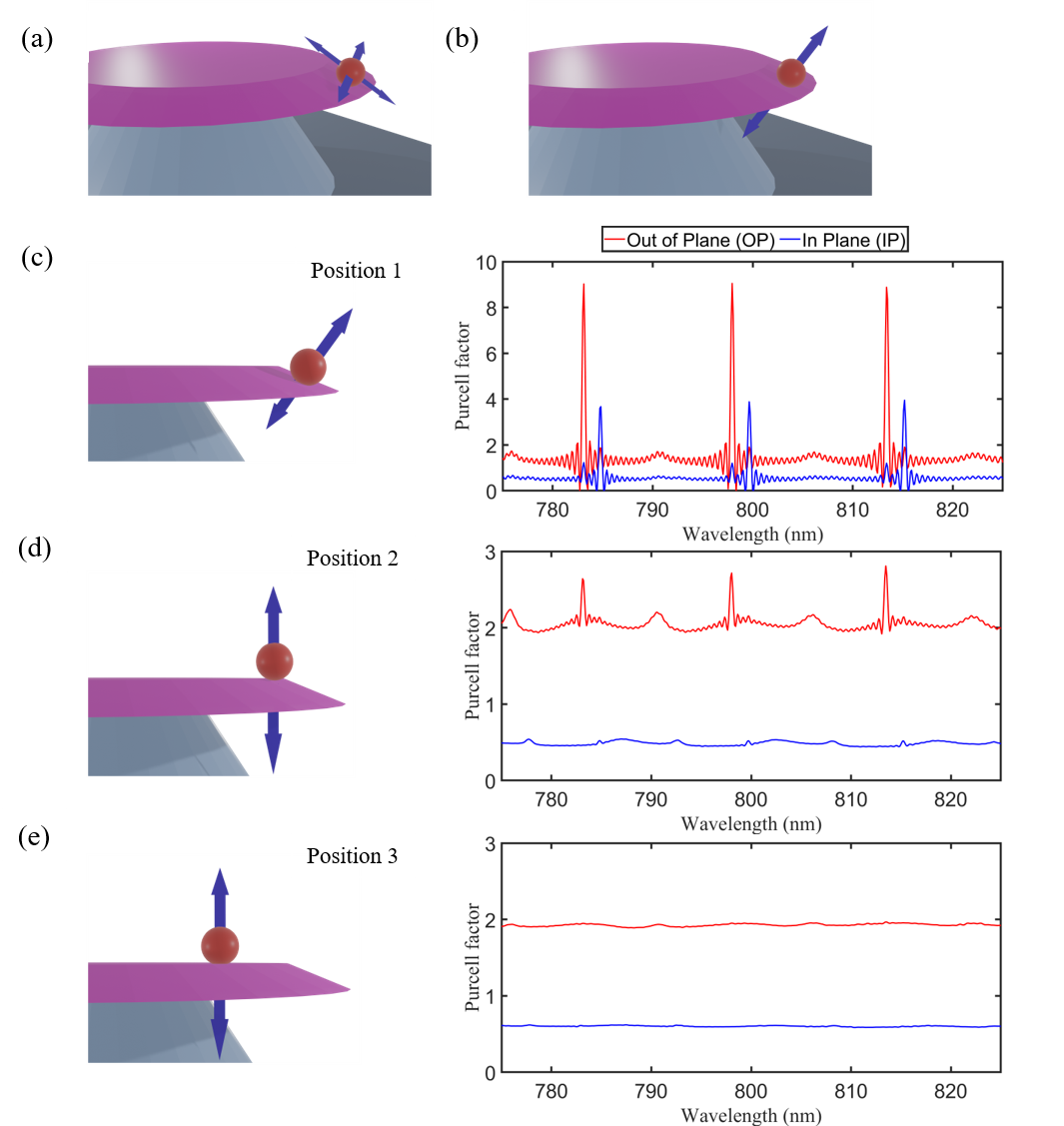}
      \caption{Depiction of (a) circularly polarised In-Plane and (b) Out-of-Plane dipole orientation in the near field of the tapered microdisk; (c)-(e) Purcell factor for IP and OP oriented dipolar emission at 800\si{nm} for positions 1,2 and 3 respectively}
    \label{fig:4}
\end{figure*}
where $\epsilon_0$ is the vacuum permittivity, $k$ is the wave number in free space and $E_\mathrm{s}$ is the scattered electric field of the emitter positioned at $r_\mathrm{e}$.
\begin{figure*}[ht]
    \centering
    \includegraphics[width=1\linewidth]{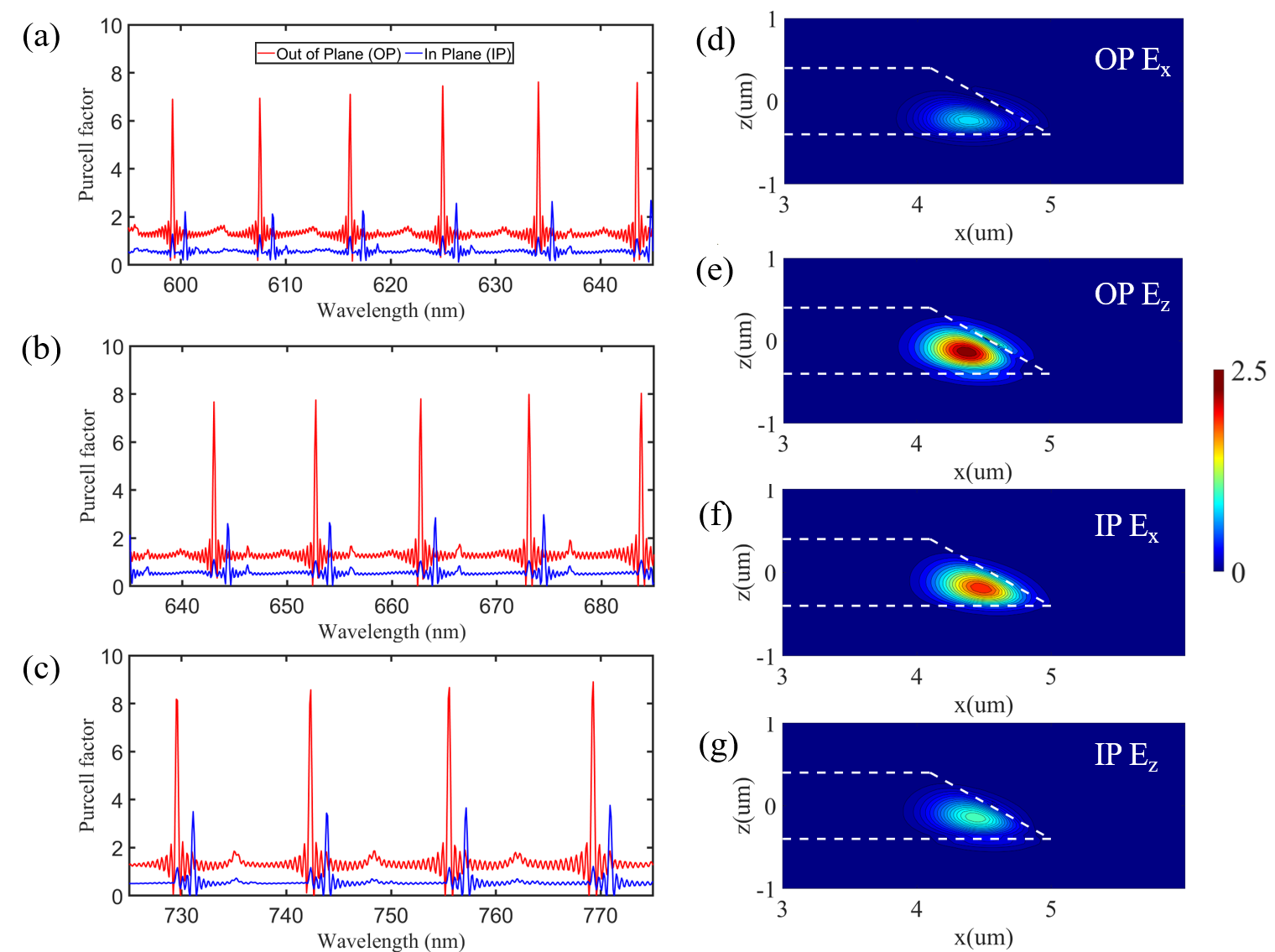}
    \caption{(a)-(c) Purcell factor of IP and OP oriented dipolar emission at 620\si{nm}, 660\si{nm} and 750\si{nm} along the taper (position 1) respectively; Electric field plots of $E_x$ and $E_z$ components at resonant peak wavelength of (d),(e) OP dipole at 742.3 nm and (f),(g) IP dipole at 743.8 nm for a dipolar emission of 750\si{nm}.}
\label{fig:5}
\end{figure*}
$PF$ can also be expressed as  the enhancement of total power emitted by an exciton either radiated to the far-field $P_{\text {rad }}$ or dissipated into the resonator or in this case that which is coupled to the WGMs of the microdisk $P_{\text {dis}}$
\begin{equation}
   PF=\left(P_{\text {rad }}+P_{\text {dis}}\right) / P_{0} 
   \label{eq:3}
\end{equation}
with $P_{0}$ being the source power. Thus the Purcell factor can be divided into radiative and non-radiative or simply the dissipated terms as
\begin{equation}
PF=PF_{\mathrm{rad}}+PF_{\text{dis}}
\end{equation}

In terms of the eigenmodes of the system $\mathbf{E}_n$, the Purcell factor is governed by the local optical density of states $\rho(\mathbf{e}_d, \mathbf{r},\omega)$ which can be written in the form\cite{Barnes2020}:
\begin{equation}
    \rho(\mathbf{e}_d, \mathbf{r},\omega)\ d\omega = \sum_n \delta(\omega-\omega_n)|\mathbf{e}_d\cdot\mathbf{E}_n(\mathbf{r})|^2 
\end{equation}
where the dipole is oriented along $\mathbf{e}_d$.
Though it is evident from the above equations that the Purcell factor is also dependent on the emitter location and orientation, we first determine the effect of $\theta$ on $PF$ with the help of integrated field strength in order to consider the average effect over the entire typical 2D TMDC flake size. 
We deduce the horizontal ($E_h$) and vertical ($E_v$) components of electric field {which are expressed by a combination of the electric field components along $x$, $y$, and $z$ axes incorporating $\theta$. The electric field is obtained using the FDE solver where the field intensity is normalized to unity. The $E_x$, $E_y$, and $E_z$ components each constitute as a subset of the electric field. The region marked between $r_1$ and $r$ exhibits significant electric field strength, defined here by $r_1 = 3\si{\um}$ for $r = 5\si{\um}$, whereas $r_2$ denotes the start of the tapering of the sidewall. The $E_h$ and $E_v$ values are integrated over the disk edge C1 with $r_1 \leq r \leq r_2 $ and the taper edge C2 ($r > r_2$) (Fig.~\ref{fig:2}(a)).}, as given in Eqs.~\ref{eq:6}--\ref{eq:7} below where $l$ represents the lines demarcated by the disk boundaries C1 and C2. 
\begin{equation}
E^2_{h_{int}}=\frac{\left(\int^{C 1}\left|E_{h}\right|^{2} d l+\int^{C 2}\left|E_{h}\right|^{2} d l\right)}{\int dl}
\label{eq:6}
\end{equation}
\begin{equation}
E^2_{v_{int}}=\frac{\left(\int^{C 1}\left|E_{v}\right|^{2} d l+\int^{C 2}\left|E_{v}\right|^{2} d l\right)}{\int dl}
\label{eq:7}
\end{equation}

{Fig.~\ref{fig:3}(d)} depicts the variation of $E_{h_{int}}$ and $E_{v_{int}}$ over a range of $\theta$ varying from $30^{\circ}$ to $90^{\circ}$.
The horizontal and vertical electric field components are clearly distinguishable when $\theta$ is above $85^{\circ}$ or less than $42^{\circ}$. Larger values of $\theta$ are difficult to be realised in fabrication since wet etching techniques inherently produce significant tapering of sidewalls. For subsequent discussion, we choose a representative $\theta = 42^{\circ}$ to study the $PF$ for dipolar emissions varying across orientation and positions. \\

We use dipolar emissions of varying wavelength equivalent to the emissions of different 2D TMDCs including WS$_2$, MoS$_2$, WSe$_2$ and MoSe$_2$ centered at $620\si{nm}$, $660\si{nm}$, $750\si{nm}$ and $800\si{nm}$. The $OP$ and circularly polarised $IP$ dipoles (Fig.~\ref{fig:4}(a),(b)) placed at different positions numbered 1 -- 3 (Fig.~\ref{fig:4}(c)-(e)) along the surface of the microdisk and the taper edge are used to excite the WGMs of the microdisk. The Purcell enhancement obtained for these configurations are numerically analysed using the transmission box of FDTD tool exploiting Eq.~\ref{eq:3}. Fig.~\ref{fig:4} shows the Purcell enhancement obtained for dipolar emission of $800 \si{nm}$ across positions 1, 2 and 3 on the microdisk. Position 1, along the taper of the microdisk, allows for a Purcell enhancement of the $OP$ dipolar emission coupled to the WGM about $5$ times greater than that of the $IP$ dipolar emission. While the dipoles on positions 2  and 3, along the top surface of the microdisk, result in weakly coupling to the WGMs of the microdisk. {The tapered microdisk clearly couples the near field of the emitter to propagating optical WGM.}

Fig.~\ref{fig:5} shows the $PF$ analysis for dipolar emissions centered at $620\si{nm}$, $660\si{nm}$ and $750\si{nm}$. While the FSR and $PF$ value are seen to increase with increase in frequency of emission, the plots reveal a similar trend being followed across different emissions where the taper allows for selective coupling of the Purcell enhanced $OP$ dipole over the $IP$ dipole. A mismatch in polarisation or frequency between the dipole and resonant cavity leads to reduction in the decay rate which may also sometimes result in $PF$ being in less than unity.The amount by which the taper prefers the $OP$ dipole over the $IP$ dipole or the ratio of Purcell enhancement obtained for $OP$ dipole to that of the $IP$ dipole ($PF_{OP}/PF_{IP}$) varies across different 2D TMDCs. This can be attributed to the frequency dependence of $PF$ as described in Eq.\ref{eq:1}. 
Though the values of ($PF_{OP}/PF_{IP}$) vary, the plots in Fig.~\ref{fig:5} for all the materials clearly depict that the tapering of disk distinguishes the in-plane and out-of-plane modes making the selective optical access of the dark and bright excitons simple and effective through their spectral signatures. This is further validated through field profiles at the resonant peaks of the WGMs $742.3\si{nm}$ and $743.8\si{nm}$ corresponding to $OP$ and $IP$ oriented dipoles for an emission at $750\si{nm}$. The $E_z$ field profile at $742.3\si{nm}$ proves the strong coupling of the WGM to the out-of-plane mode while the $E_x$  profile at $743.8\si{nm}$ illustrates the dominance of in-plane coupled WGM. The relative field magnitudes also depict the preference of out-of-plane mode over the in-plane mode.

\section{Conclusion}
Dark excitons impact the exciton dynamics significantly in 2D TMDCs\cite{PhysRevMaterials.2.014002}. They possess a great potential for light storage, communication and information processing applications owing to their longer radiative lifetime\cite{PhysRevX.5.011009}. We have shown how their optical accessibility can be enhanced and tailored by coupling of 2D TMDCs with tapered microdisks. We showed that the whispering gallery modes in a silica microdisk can selectively couple the Purcell enhanced dark and bright excitons when the 2D TMDC is positioned along the tapered sidewall. Thus, tailoring of the excitonic light emission in 2D TMDCs can be achieved by coupling them with morphologically engineered WGM resonators. In particular, we have shown that engineering of the taper angle in a microdisk is a versatile route to achieving such a Purcell anisotropy for four representative 2D TMDCs. Our work should open up new avenues to explore this microdisk taper platform for studying exotic phenomena such as anisotropic vacuum induced excitonic valley coherence\cite{PhysRevLett.121.116102,PhysRevB.102.045416} and strong coupling\cite{Schneider2018,PhysRevB.98.161113} of dark excitons with WGMs and should also serve as a convenient probe for other exotic polaritons in 2D semiconductors, in particular interlayer excitons in heterostructures\cite{Jiang2021,Sohoni2020}.
 
\emph{Acknowledgement.} A.K. acknowledges funding from the Department of Science and Technology via the grants: SB/S2/RJN-110/2017, ECR/2018/001485 and DST/NM/NS-2018/49.

\bibliography{references.bib}

\begin{thebibliography}{44}%
\makeatletter
\providecommand \@ifxundefined [1]{%
 \@ifx{#1\undefined}
}%
\providecommand \@ifnum [1]{%
 \ifnum #1\expandafter \@firstoftwo
 \else \expandafter \@secondoftwo
 \fi
}%
\providecommand \@ifx [1]{%
 \ifx #1\expandafter \@firstoftwo
 \else \expandafter \@secondoftwo
 \fi
}%
\providecommand \natexlab [1]{#1}%
\providecommand \enquote  [1]{``#1''}%
\providecommand \bibnamefont  [1]{#1}%
\providecommand \bibfnamefont [1]{#1}%
\providecommand \citenamefont [1]{#1}%
\providecommand \href@noop [0]{\@secondoftwo}%
\providecommand \href [0]{\begingroup \@sanitize@url \@href}%
\providecommand \@href[1]{\@@startlink{#1}\@@href}%
\providecommand \@@href[1]{\endgroup#1\@@endlink}%
\providecommand \@sanitize@url [0]{\catcode `\\12\catcode `\$12\catcode
  `\&12\catcode `\#12\catcode `\^12\catcode `\_12\catcode `\%12\relax}%
\providecommand \@@startlink[1]{}%
\providecommand \@@endlink[0]{}%
\providecommand \url  [0]{\begingroup\@sanitize@url \@url }%
\providecommand \@url [1]{\endgroup\@href {#1}{\urlprefix }}%
\providecommand \urlprefix  [0]{URL }%
\providecommand \Eprint [0]{\href }%
\providecommand \doibase [0]{https://doi.org/}%
\providecommand \selectlanguage [0]{\@gobble}%
\providecommand \bibinfo  [0]{\@secondoftwo}%
\providecommand \bibfield  [0]{\@secondoftwo}%
\providecommand \translation [1]{[#1]}%
\providecommand \BibitemOpen [0]{}%
\providecommand \bibitemStop [0]{}%
\providecommand \bibitemNoStop [0]{.\EOS\space}%
\providecommand \EOS [0]{\spacefactor3000\relax}%
\providecommand \BibitemShut  [1]{\csname bibitem#1\endcsname}%
\let\auto@bib@innerbib\@empty
\bibitem [{\citenamefont {Khan}\ \emph {et~al.}(2020)\citenamefont {Khan},
  \citenamefont {Tareen}, \citenamefont {Aslam}, \citenamefont {Wang},
  \citenamefont {Zhang}, \citenamefont {Mahmood}, \citenamefont {Ouyang},
  \citenamefont {Zhang},\ and\ \citenamefont {Guo}}]{Khan2020a}%
  \BibitemOpen
  \bibfield  {author} {\bibinfo {author} {\bibfnamefont {K.}~\bibnamefont
  {Khan}}, \bibinfo {author} {\bibfnamefont {A.~K.}\ \bibnamefont {Tareen}},
  \bibinfo {author} {\bibfnamefont {M.}~\bibnamefont {Aslam}}, \bibinfo
  {author} {\bibfnamefont {R.}~\bibnamefont {Wang}}, \bibinfo {author}
  {\bibfnamefont {Y.}~\bibnamefont {Zhang}}, \bibinfo {author} {\bibfnamefont
  {A.}~\bibnamefont {Mahmood}}, \bibinfo {author} {\bibfnamefont
  {Z.}~\bibnamefont {Ouyang}}, \bibinfo {author} {\bibfnamefont
  {H.}~\bibnamefont {Zhang}},\ and\ \bibinfo {author} {\bibfnamefont
  {Z.}~\bibnamefont {Guo}},\ }\bibfield  {title} {\bibinfo {title} {Recent
  developments in emerging two-dimensional materials and their applications},\
  }\href {https://doi.org/10.1039/c9tc04187g} {\bibfield  {journal} {\bibinfo
  {journal} {Journal of Materials Chemistry C}\ }\textbf {\bibinfo {volume}
  {8}},\ \bibinfo {pages} {387} (\bibinfo {year} {2020})}\BibitemShut {NoStop}%
\bibitem [{\citenamefont {Glavin}\ \emph {et~al.}(2020)\citenamefont {Glavin},
  \citenamefont {Rao}, \citenamefont {Varshney}, \citenamefont {Bianco},
  \citenamefont {Apte}, \citenamefont {Roy}, \citenamefont {Ringe},\ and\
  \citenamefont {Ajayan}}]{Glavin2020a}%
  \BibitemOpen
  \bibfield  {author} {\bibinfo {author} {\bibfnamefont {N.~R.}\ \bibnamefont
  {Glavin}}, \bibinfo {author} {\bibfnamefont {R.}~\bibnamefont {Rao}},
  \bibinfo {author} {\bibfnamefont {V.}~\bibnamefont {Varshney}}, \bibinfo
  {author} {\bibfnamefont {E.}~\bibnamefont {Bianco}}, \bibinfo {author}
  {\bibfnamefont {A.}~\bibnamefont {Apte}}, \bibinfo {author} {\bibfnamefont
  {A.}~\bibnamefont {Roy}}, \bibinfo {author} {\bibfnamefont {E.}~\bibnamefont
  {Ringe}},\ and\ \bibinfo {author} {\bibfnamefont {P.~M.}\ \bibnamefont
  {Ajayan}},\ }\bibfield  {title} {\bibinfo {title} {{Emerging Applications of
  Elemental 2D Materials}},\ }\href {https://doi.org/10.1002/adma.201904302}
  {\bibfield  {journal} {\bibinfo  {journal} {Advanced Materials}\ }\textbf
  {\bibinfo {volume} {32}},\ \bibinfo {pages} {1} (\bibinfo {year}
  {2020})}\BibitemShut {NoStop}%
\bibitem [{\citenamefont {Krasnok}\ \emph {et~al.}(2018)\citenamefont
  {Krasnok}, \citenamefont {Lepeshov},\ and\ \citenamefont
  {Al{\'{u}}}}]{Rasnok2014NanophotonicsDichalcogenides}%
  \BibitemOpen
  \bibfield  {author} {\bibinfo {author} {\bibfnamefont {A.}~\bibnamefont
  {Krasnok}}, \bibinfo {author} {\bibfnamefont {S.}~\bibnamefont {Lepeshov}},\
  and\ \bibinfo {author} {\bibfnamefont {A.}~\bibnamefont {Al{\'{u}}}},\
  }\bibfield  {title} {\bibinfo {title} {Nanophotonics with 2d transition metal
  dichalcogenides [invited]},\ }\href {https://doi.org/10.1364/oe.26.015972}
  {\bibfield  {journal} {\bibinfo  {journal} {Optics Express}\ }\textbf
  {\bibinfo {volume} {26}},\ \bibinfo {pages} {15972} (\bibinfo {year}
  {2018})}\BibitemShut {NoStop}%
\bibitem [{\citenamefont {Mueller}\ and\ \citenamefont
  {Malic}(2018)}]{Mueller2018ExcitonSemiconductors}%
  \BibitemOpen
  \bibfield  {author} {\bibinfo {author} {\bibfnamefont {T.}~\bibnamefont
  {Mueller}}\ and\ \bibinfo {author} {\bibfnamefont {E.}~\bibnamefont
  {Malic}},\ }\bibfield  {title} {\bibinfo {title} {{Exciton physics and device
  application of two-dimensional transition metal dichalcogenide
  semiconductors}},\ }\href {https://doi.org/10.1038/s41699-018-0074-2}
  {\bibfield  {journal} {\bibinfo  {journal} {npj 2D Materials and
  Applications}\ ,\ \bibinfo {pages} {1}} (\bibinfo {year} {2018})}\BibitemShut
  {NoStop}%
\bibitem [{\citenamefont {Yuan}\ \emph {et~al.}(2017)\citenamefont {Yuan},
  \citenamefont {Wang}, \citenamefont {Zhu}, \citenamefont {Zhou},\ and\
  \citenamefont {Huang}}]{Yuan2017ExcitonSemiconductors}%
  \BibitemOpen
  \bibfield  {author} {\bibinfo {author} {\bibfnamefont {L.}~\bibnamefont
  {Yuan}}, \bibinfo {author} {\bibfnamefont {T.}~\bibnamefont {Wang}}, \bibinfo
  {author} {\bibfnamefont {T.}~\bibnamefont {Zhu}}, \bibinfo {author}
  {\bibfnamefont {M.}~\bibnamefont {Zhou}},\ and\ \bibinfo {author}
  {\bibfnamefont {L.}~\bibnamefont {Huang}},\ }\bibfield  {title} {\bibinfo
  {title} {Exciton dynamics, transport, and annihilation in atomically thin
  two-dimensional semiconductors},\ }\href
  {https://doi.org/10.1021/acs.jpclett.7b00885} {\bibfield  {journal} {\bibinfo
   {journal} {The Journal of Physical Chemistry Letters}\ }\textbf {\bibinfo
  {volume} {8}},\ \bibinfo {pages} {3371} (\bibinfo {year} {2017})}\BibitemShut
  {NoStop}%
\bibitem [{\citenamefont {Moody}\ \emph {et~al.}(2016)\citenamefont {Moody},
  \citenamefont {Schaibley},\ and\ \citenamefont {Xu}}]{Oody2016Exciton}%
  \BibitemOpen
  \bibfield  {author} {\bibinfo {author} {\bibfnamefont {G.}~\bibnamefont
  {Moody}}, \bibinfo {author} {\bibfnamefont {J.}~\bibnamefont {Schaibley}},\
  and\ \bibinfo {author} {\bibfnamefont {X.}~\bibnamefont {Xu}},\ }\bibfield
  {title} {\bibinfo {title} {Exciton dynamics in monolayer transition metal
  dichalcogenides},\ }\href {https://doi.org/10.1364/josab.33.000c39}
  {\bibfield  {journal} {\bibinfo  {journal} {Journal of the Optical Society of
  America B}\ }\textbf {\bibinfo {volume} {33}},\ \bibinfo {pages} {C39}
  (\bibinfo {year} {2016})}\BibitemShut {NoStop}%
\bibitem [{\citenamefont {Jiang}\ \emph
  {et~al.}(2021{\natexlab{a}})\citenamefont {Jiang}, \citenamefont {Zheng},
  \citenamefont {Lan}, \citenamefont {Saidi}, \citenamefont {Ren},\ and\
  \citenamefont {Zhao}}]{Jiang2021Real-timeDichalcogenide}%
  \BibitemOpen
  \bibfield  {author} {\bibinfo {author} {\bibfnamefont {X.}~\bibnamefont
  {Jiang}}, \bibinfo {author} {\bibfnamefont {Q.}~\bibnamefont {Zheng}},
  \bibinfo {author} {\bibfnamefont {Z.}~\bibnamefont {Lan}}, \bibinfo {author}
  {\bibfnamefont {W.~A.}\ \bibnamefont {Saidi}}, \bibinfo {author}
  {\bibfnamefont {X.}~\bibnamefont {Ren}},\ and\ \bibinfo {author}
  {\bibfnamefont {J.}~\bibnamefont {Zhao}},\ }\bibfield  {title} {\bibinfo
  {title} {Real-time {GW}-{BSE} investigations on spin-valley exciton dynamics
  in monolayer transition metal dichalcogenide},\ }\href
  {https://doi.org/10.1126/sciadv.abf3759} {\bibfield  {journal} {\bibinfo
  {journal} {Science Advances}\ }\textbf {\bibinfo {volume} {7}},\ \bibinfo
  {pages} {eabf3759} (\bibinfo {year} {2021}{\natexlab{a}})}\BibitemShut
  {NoStop}%
\bibitem [{\citenamefont {Selig}\ \emph {et~al.}(2018)\citenamefont {Selig},
  \citenamefont {Bergh\"{a}user}, \citenamefont {Richter}, \citenamefont
  {Bratschitsch}, \citenamefont {Knorr},\ and\ \citenamefont
  {Malic}}]{Selig2018AndDichalcogenides}%
  \BibitemOpen
  \bibfield  {author} {\bibinfo {author} {\bibfnamefont {M.}~\bibnamefont
  {Selig}}, \bibinfo {author} {\bibfnamefont {G.}~\bibnamefont
  {Bergh\"{a}user}}, \bibinfo {author} {\bibfnamefont {M.}~\bibnamefont
  {Richter}}, \bibinfo {author} {\bibfnamefont {R.}~\bibnamefont
  {Bratschitsch}}, \bibinfo {author} {\bibfnamefont {A.}~\bibnamefont
  {Knorr}},\ and\ \bibinfo {author} {\bibfnamefont {E.}~\bibnamefont {Malic}},\
  }\bibfield  {title} {\bibinfo {title} {Dark and bright exciton formation,
  thermalization, and photoluminescence in monolayer transition metal
  dichalcogenides},\ }\href {https://doi.org/10.1088/2053-1583/aabea3}
  {\bibfield  {journal} {\bibinfo  {journal} {2D Materials}\ }\textbf {\bibinfo
  {volume} {5}},\ \bibinfo {pages} {035017} (\bibinfo {year}
  {2018})}\BibitemShut {NoStop}%
\bibitem [{\citenamefont {Loh}(2017)}]{LohBrighteningArrays}%
  \BibitemOpen
  \bibfield  {author} {\bibinfo {author} {\bibfnamefont {K.~P.}\ \bibnamefont
  {Loh}},\ }\bibfield  {title} {\bibinfo {title} {Brightening the dark
  excitons},\ }\href {https://doi.org/10.1038/nnano.2017.130} {\bibfield
  {journal} {\bibinfo  {journal} {Nature Nanotechnology}\ }\textbf {\bibinfo
  {volume} {12}},\ \bibinfo {pages} {837} (\bibinfo {year} {2017})}\BibitemShut
  {NoStop}%
\bibitem [{\citenamefont {Na}\ and\ \citenamefont
  {Ye}(2020)}]{Meng2020ProbingExciton}%
  \BibitemOpen
  \bibfield  {author} {\bibinfo {author} {\bibfnamefont {M.~X.}\ \bibnamefont
  {Na}}\ and\ \bibinfo {author} {\bibfnamefont {Z.}~\bibnamefont {Ye}},\
  }\bibfield  {title} {\bibinfo {title} {Probing the dark side of the
  exciton},\ }\href {https://doi.org/10.1126/science.abf0371} {\bibfield
  {journal} {\bibinfo  {journal} {Science}\ }\textbf {\bibinfo {volume}
  {370}},\ \bibinfo {pages} {1166} (\bibinfo {year} {2020})}\BibitemShut
  {NoStop}%
\bibitem [{\citenamefont {Lu}\ \emph {et~al.}(2019)\citenamefont {Lu},
  \citenamefont {Rhodes}, \citenamefont {Li}, \citenamefont {Tuan},
  \citenamefont {Jiang}, \citenamefont {Ludwig}, \citenamefont {Jiang},
  \citenamefont {Lian}, \citenamefont {Shi}, \citenamefont {Hone},
  \citenamefont {Dery},\ and\ \citenamefont {Smirnov}}]{Lu2019}%
  \BibitemOpen
  \bibfield  {author} {\bibinfo {author} {\bibfnamefont {Z.}~\bibnamefont
  {Lu}}, \bibinfo {author} {\bibfnamefont {D.}~\bibnamefont {Rhodes}}, \bibinfo
  {author} {\bibfnamefont {Z.}~\bibnamefont {Li}}, \bibinfo {author}
  {\bibfnamefont {D.~V.}\ \bibnamefont {Tuan}}, \bibinfo {author}
  {\bibfnamefont {Y.}~\bibnamefont {Jiang}}, \bibinfo {author} {\bibfnamefont
  {J.}~\bibnamefont {Ludwig}}, \bibinfo {author} {\bibfnamefont
  {Z.}~\bibnamefont {Jiang}}, \bibinfo {author} {\bibfnamefont
  {Z.}~\bibnamefont {Lian}}, \bibinfo {author} {\bibfnamefont {S.-F.}\
  \bibnamefont {Shi}}, \bibinfo {author} {\bibfnamefont {J.}~\bibnamefont
  {Hone}}, \bibinfo {author} {\bibfnamefont {H.}~\bibnamefont {Dery}},\ and\
  \bibinfo {author} {\bibfnamefont {D.}~\bibnamefont {Smirnov}},\ }\bibfield
  {title} {\bibinfo {title} {Magnetic field mixing and splitting of bright and
  dark excitons in monolayer {MoSe} 2},\ }\href
  {https://doi.org/10.1088/2053-1583/ab5614} {\bibfield  {journal} {\bibinfo
  {journal} {2D Materials}\ }\textbf {\bibinfo {volume} {7}},\ \bibinfo {pages}
  {015017} (\bibinfo {year} {2019})}\BibitemShut {NoStop}%
\bibitem [{\citenamefont {Molas}\ \emph {et~al.}(2019)\citenamefont {Molas},
  \citenamefont {Slobodeniuk}, \citenamefont {Kazimierczuk}, \citenamefont
  {Nogajewski}, \citenamefont {Bartos}, \citenamefont {Kapu\ifmmode
  \acute{s}\else \'{s}\fi{}ci\ifmmode~\acute{n}\else \'{n}\fi{}ski},
  \citenamefont {Oreszczuk}, \citenamefont {Watanabe}, \citenamefont
  {Taniguchi}, \citenamefont {Faugeras}, \citenamefont {Kossacki},
  \citenamefont {Basko},\ and\ \citenamefont
  {Potemski}}]{PhysRevLett.123.096803}%
  \BibitemOpen
  \bibfield  {author} {\bibinfo {author} {\bibfnamefont {M.~R.}\ \bibnamefont
  {Molas}}, \bibinfo {author} {\bibfnamefont {A.~O.}\ \bibnamefont
  {Slobodeniuk}}, \bibinfo {author} {\bibfnamefont {T.}~\bibnamefont
  {Kazimierczuk}}, \bibinfo {author} {\bibfnamefont {K.}~\bibnamefont
  {Nogajewski}}, \bibinfo {author} {\bibfnamefont {M.}~\bibnamefont {Bartos}},
  \bibinfo {author} {\bibfnamefont {P.}~\bibnamefont {Kapu\ifmmode
  \acute{s}\else \'{s}\fi{}ci\ifmmode~\acute{n}\else \'{n}\fi{}ski}}, \bibinfo
  {author} {\bibfnamefont {K.}~\bibnamefont {Oreszczuk}}, \bibinfo {author}
  {\bibfnamefont {K.}~\bibnamefont {Watanabe}}, \bibinfo {author}
  {\bibfnamefont {T.}~\bibnamefont {Taniguchi}}, \bibinfo {author}
  {\bibfnamefont {C.}~\bibnamefont {Faugeras}}, \bibinfo {author}
  {\bibfnamefont {P.}~\bibnamefont {Kossacki}}, \bibinfo {author}
  {\bibfnamefont {D.~M.}\ \bibnamefont {Basko}},\ and\ \bibinfo {author}
  {\bibfnamefont {M.}~\bibnamefont {Potemski}},\ }\bibfield  {title} {\bibinfo
  {title} {Probing and manipulating valley coherence of dark excitons in
  monolayer ${\mathrm{wse}}_{2}$},\ }\href
  {https://doi.org/10.1103/PhysRevLett.123.096803} {\bibfield  {journal}
  {\bibinfo  {journal} {Phys. Rev. Lett.}\ }\textbf {\bibinfo {volume} {123}},\
  \bibinfo {pages} {096803} (\bibinfo {year} {2019})}\BibitemShut {NoStop}%
\bibitem [{\citenamefont {Zhang}\ \emph {et~al.}(2017)\citenamefont {Zhang},
  \citenamefont {Cao}, \citenamefont {Lu}, \citenamefont {Lin}, \citenamefont
  {Zhang}, \citenamefont {Wang}, \citenamefont {Li}, \citenamefont {Hone},
  \citenamefont {Robinson}, \citenamefont {Smirnov}, \citenamefont {Louie},\
  and\ \citenamefont {Heinz}}]{Zhang2017}%
  \BibitemOpen
  \bibfield  {author} {\bibinfo {author} {\bibfnamefont {X.-X.}\ \bibnamefont
  {Zhang}}, \bibinfo {author} {\bibfnamefont {T.}~\bibnamefont {Cao}}, \bibinfo
  {author} {\bibfnamefont {Z.}~\bibnamefont {Lu}}, \bibinfo {author}
  {\bibfnamefont {Y.-C.}\ \bibnamefont {Lin}}, \bibinfo {author} {\bibfnamefont
  {F.}~\bibnamefont {Zhang}}, \bibinfo {author} {\bibfnamefont
  {Y.}~\bibnamefont {Wang}}, \bibinfo {author} {\bibfnamefont {Z.}~\bibnamefont
  {Li}}, \bibinfo {author} {\bibfnamefont {J.~C.}\ \bibnamefont {Hone}},
  \bibinfo {author} {\bibfnamefont {J.~A.}\ \bibnamefont {Robinson}}, \bibinfo
  {author} {\bibfnamefont {D.}~\bibnamefont {Smirnov}}, \bibinfo {author}
  {\bibfnamefont {S.~G.}\ \bibnamefont {Louie}},\ and\ \bibinfo {author}
  {\bibfnamefont {T.~F.}\ \bibnamefont {Heinz}},\ }\bibfield  {title} {\bibinfo
  {title} {Magnetic brightening and control of dark excitons in monolayer
  {WSe}2},\ }\href {https://doi.org/10.1038/nnano.2017.105} {\bibfield
  {journal} {\bibinfo  {journal} {Nature Nanotechnology}\ }\textbf {\bibinfo
  {volume} {12}},\ \bibinfo {pages} {883} (\bibinfo {year} {2017})}\BibitemShut
  {NoStop}%
\bibitem [{\citenamefont {Molas}\ \emph {et~al.}(2017)\citenamefont {Molas},
  \citenamefont {Faugeras}, \citenamefont {Slobodeniuk}, \citenamefont
  {Nogajewski}, \citenamefont {Bartos}, \citenamefont {Basko},\ and\
  \citenamefont {Potemski}}]{Molas2017}%
  \BibitemOpen
  \bibfield  {author} {\bibinfo {author} {\bibfnamefont {M.~R.}\ \bibnamefont
  {Molas}}, \bibinfo {author} {\bibfnamefont {C.}~\bibnamefont {Faugeras}},
  \bibinfo {author} {\bibfnamefont {A.~O.}\ \bibnamefont {Slobodeniuk}},
  \bibinfo {author} {\bibfnamefont {K.}~\bibnamefont {Nogajewski}}, \bibinfo
  {author} {\bibfnamefont {M.}~\bibnamefont {Bartos}}, \bibinfo {author}
  {\bibfnamefont {D.~M.}\ \bibnamefont {Basko}},\ and\ \bibinfo {author}
  {\bibfnamefont {M.}~\bibnamefont {Potemski}},\ }\bibfield  {title} {\bibinfo
  {title} {Brightening of dark excitons in monolayers of semiconducting
  transition metal dichalcogenides},\ }\href
  {https://doi.org/10.1088/2053-1583/aa5521} {\bibfield  {journal} {\bibinfo
  {journal} {2D Materials}\ }\textbf {\bibinfo {volume} {4}},\ \bibinfo {pages}
  {021003} (\bibinfo {year} {2017})}\BibitemShut {NoStop}%
\bibitem [{\citenamefont {Robert}\ \emph {et~al.}(2017)\citenamefont {Robert},
  \citenamefont {Amand}, \citenamefont {Cadiz}, \citenamefont {Lagarde},
  \citenamefont {Courtade}, \citenamefont {Manca}, \citenamefont {Taniguchi},
  \citenamefont {Watanabe}, \citenamefont {Urbaszek},\ and\ \citenamefont
  {Marie}}]{PhysRevB.96.155423}%
  \BibitemOpen
  \bibfield  {author} {\bibinfo {author} {\bibfnamefont {C.}~\bibnamefont
  {Robert}}, \bibinfo {author} {\bibfnamefont {T.}~\bibnamefont {Amand}},
  \bibinfo {author} {\bibfnamefont {F.}~\bibnamefont {Cadiz}}, \bibinfo
  {author} {\bibfnamefont {D.}~\bibnamefont {Lagarde}}, \bibinfo {author}
  {\bibfnamefont {E.}~\bibnamefont {Courtade}}, \bibinfo {author}
  {\bibfnamefont {M.}~\bibnamefont {Manca}}, \bibinfo {author} {\bibfnamefont
  {T.}~\bibnamefont {Taniguchi}}, \bibinfo {author} {\bibfnamefont
  {K.}~\bibnamefont {Watanabe}}, \bibinfo {author} {\bibfnamefont
  {B.}~\bibnamefont {Urbaszek}},\ and\ \bibinfo {author} {\bibfnamefont
  {X.}~\bibnamefont {Marie}},\ }\bibfield  {title} {\bibinfo {title} {Fine
  structure and lifetime of dark excitons in transition metal dichalcogenide
  monolayers},\ }\href {https://doi.org/10.1103/PhysRevB.96.155423} {\bibfield
  {journal} {\bibinfo  {journal} {Phys. Rev. B}\ }\textbf {\bibinfo {volume}
  {96}},\ \bibinfo {pages} {155423} (\bibinfo {year} {2017})}\BibitemShut
  {NoStop}%
\bibitem [{\citenamefont {Robert}\ \emph {et~al.}(2020)\citenamefont {Robert},
  \citenamefont {Han}, \citenamefont {Kapuscinski}, \citenamefont {Delhomme},
  \citenamefont {Faugeras}, \citenamefont {Amand}, \citenamefont {Molas},
  \citenamefont {Bartos}, \citenamefont {Watanabe}, \citenamefont {Taniguchi},
  \citenamefont {Urbaszek}, \citenamefont {Potemski},\ and\ \citenamefont
  {Marie}}]{RobertMeasurementExcitons}%
  \BibitemOpen
  \bibfield  {author} {\bibinfo {author} {\bibfnamefont {C.}~\bibnamefont
  {Robert}}, \bibinfo {author} {\bibfnamefont {B.}~\bibnamefont {Han}},
  \bibinfo {author} {\bibfnamefont {P.}~\bibnamefont {Kapuscinski}}, \bibinfo
  {author} {\bibfnamefont {A.}~\bibnamefont {Delhomme}}, \bibinfo {author}
  {\bibfnamefont {C.}~\bibnamefont {Faugeras}}, \bibinfo {author}
  {\bibfnamefont {T.}~\bibnamefont {Amand}}, \bibinfo {author} {\bibfnamefont
  {M.~R.}\ \bibnamefont {Molas}}, \bibinfo {author} {\bibfnamefont
  {M.}~\bibnamefont {Bartos}}, \bibinfo {author} {\bibfnamefont
  {K.}~\bibnamefont {Watanabe}}, \bibinfo {author} {\bibfnamefont
  {T.}~\bibnamefont {Taniguchi}}, \bibinfo {author} {\bibfnamefont
  {B.}~\bibnamefont {Urbaszek}}, \bibinfo {author} {\bibfnamefont
  {M.}~\bibnamefont {Potemski}},\ and\ \bibinfo {author} {\bibfnamefont
  {X.}~\bibnamefont {Marie}},\ }\bibfield  {title} {\bibinfo {title}
  {Measurement of the spin-forbidden dark excitons in {MoS}2 and {MoSe}2
  monolayers},\ }\bibfield  {journal} {\bibinfo  {journal} {Nature
  Communications}\ }\textbf {\bibinfo {volume} {11}},\ \href
  {https://doi.org/10.1038/s41467-020-17608-4} {10.1038/s41467-020-17608-4}
  (\bibinfo {year} {2020})\BibitemShut {NoStop}%
\bibitem [{\citenamefont {Zhou}\ \emph {et~al.}(2017)\citenamefont {Zhou},
  \citenamefont {Scuri}, \citenamefont {Wild}, \citenamefont {High},
  \citenamefont {Dibos}, \citenamefont {Jauregui}, \citenamefont {Shu},
  \citenamefont {Greve}, \citenamefont {Pistunova}, \citenamefont {Joe},
  \citenamefont {Taniguchi}, \citenamefont {Watanabe}, \citenamefont {Kim},
  \citenamefont {Lukin},\ and\ \citenamefont {Park}}]{Zhou2017}%
  \BibitemOpen
  \bibfield  {author} {\bibinfo {author} {\bibfnamefont {Y.}~\bibnamefont
  {Zhou}}, \bibinfo {author} {\bibfnamefont {G.}~\bibnamefont {Scuri}},
  \bibinfo {author} {\bibfnamefont {D.~S.}\ \bibnamefont {Wild}}, \bibinfo
  {author} {\bibfnamefont {A.~A.}\ \bibnamefont {High}}, \bibinfo {author}
  {\bibfnamefont {A.}~\bibnamefont {Dibos}}, \bibinfo {author} {\bibfnamefont
  {L.~A.}\ \bibnamefont {Jauregui}}, \bibinfo {author} {\bibfnamefont
  {C.}~\bibnamefont {Shu}}, \bibinfo {author} {\bibfnamefont {K.~D.}\
  \bibnamefont {Greve}}, \bibinfo {author} {\bibfnamefont {K.}~\bibnamefont
  {Pistunova}}, \bibinfo {author} {\bibfnamefont {A.~Y.}\ \bibnamefont {Joe}},
  \bibinfo {author} {\bibfnamefont {T.}~\bibnamefont {Taniguchi}}, \bibinfo
  {author} {\bibfnamefont {K.}~\bibnamefont {Watanabe}}, \bibinfo {author}
  {\bibfnamefont {P.}~\bibnamefont {Kim}}, \bibinfo {author} {\bibfnamefont
  {M.~D.}\ \bibnamefont {Lukin}},\ and\ \bibinfo {author} {\bibfnamefont
  {H.}~\bibnamefont {Park}},\ }\bibfield  {title} {\bibinfo {title} {Probing
  dark excitons in atomically thin semiconductors via near-field coupling to
  surface plasmon polaritons},\ }\href {https://doi.org/10.1038/nnano.2017.106}
  {\bibfield  {journal} {\bibinfo  {journal} {Nature Nanotechnology}\ }\textbf
  {\bibinfo {volume} {12}},\ \bibinfo {pages} {856} (\bibinfo {year}
  {2017})}\BibitemShut {NoStop}%
\bibitem [{\citenamefont {Park}\ \emph {et~al.}(2017)\citenamefont {Park},
  \citenamefont {Jiang}, \citenamefont {Clark}, \citenamefont {Xu},\ and\
  \citenamefont {Raschke}}]{Park2017}%
  \BibitemOpen
  \bibfield  {author} {\bibinfo {author} {\bibfnamefont {K.-D.}\ \bibnamefont
  {Park}}, \bibinfo {author} {\bibfnamefont {T.}~\bibnamefont {Jiang}},
  \bibinfo {author} {\bibfnamefont {G.}~\bibnamefont {Clark}}, \bibinfo
  {author} {\bibfnamefont {X.}~\bibnamefont {Xu}},\ and\ \bibinfo {author}
  {\bibfnamefont {M.~B.}\ \bibnamefont {Raschke}},\ }\bibfield  {title}
  {\bibinfo {title} {Radiative control of dark excitons at room temperature by
  nano-optical antenna-tip purcell effect},\ }\href
  {https://doi.org/10.1038/s41565-017-0003-0} {\bibfield  {journal} {\bibinfo
  {journal} {Nature Nanotechnology}\ }\textbf {\bibinfo {volume} {13}},\
  \bibinfo {pages} {59} (\bibinfo {year} {2017})}\BibitemShut {NoStop}%
\bibitem [{\citenamefont {Wang}\ \emph {et~al.}(2017)\citenamefont {Wang},
  \citenamefont {Robert}, \citenamefont {Glazov}, \citenamefont {Cadiz},
  \citenamefont {Courtade}, \citenamefont {Amand}, \citenamefont {Lagarde},
  \citenamefont {Taniguchi}, \citenamefont {Watanabe}, \citenamefont
  {Urbaszek},\ and\ \citenamefont {Marie}}]{PhysRevLett.119.047401}%
  \BibitemOpen
  \bibfield  {author} {\bibinfo {author} {\bibfnamefont {G.}~\bibnamefont
  {Wang}}, \bibinfo {author} {\bibfnamefont {C.}~\bibnamefont {Robert}},
  \bibinfo {author} {\bibfnamefont {M.~M.}\ \bibnamefont {Glazov}}, \bibinfo
  {author} {\bibfnamefont {F.}~\bibnamefont {Cadiz}}, \bibinfo {author}
  {\bibfnamefont {E.}~\bibnamefont {Courtade}}, \bibinfo {author}
  {\bibfnamefont {T.}~\bibnamefont {Amand}}, \bibinfo {author} {\bibfnamefont
  {D.}~\bibnamefont {Lagarde}}, \bibinfo {author} {\bibfnamefont
  {T.}~\bibnamefont {Taniguchi}}, \bibinfo {author} {\bibfnamefont
  {K.}~\bibnamefont {Watanabe}}, \bibinfo {author} {\bibfnamefont
  {B.}~\bibnamefont {Urbaszek}},\ and\ \bibinfo {author} {\bibfnamefont
  {X.}~\bibnamefont {Marie}},\ }\bibfield  {title} {\bibinfo {title} {In-plane
  propagation of light in transition metal dichalcogenide monolayers: Optical
  selection rules},\ }\href {https://doi.org/10.1103/PhysRevLett.119.047401}
  {\bibfield  {journal} {\bibinfo  {journal} {Phys. Rev. Lett.}\ }\textbf
  {\bibinfo {volume} {119}},\ \bibinfo {pages} {047401} (\bibinfo {year}
  {2017})}\BibitemShut {NoStop}%
\bibitem [{\citenamefont {Schneider}\ \emph {et~al.}(2020)\citenamefont
  {Schneider}, \citenamefont {Esdaille}, \citenamefont {Rhodes}, \citenamefont
  {Barmak}, \citenamefont {Hone},\ and\ \citenamefont
  {Rahimi-Iman}}]{Schneider2020}%
  \BibitemOpen
  \bibfield  {author} {\bibinfo {author} {\bibfnamefont {L.~M.}\ \bibnamefont
  {Schneider}}, \bibinfo {author} {\bibfnamefont {S.~S.}\ \bibnamefont
  {Esdaille}}, \bibinfo {author} {\bibfnamefont {D.~A.}\ \bibnamefont
  {Rhodes}}, \bibinfo {author} {\bibfnamefont {K.}~\bibnamefont {Barmak}},
  \bibinfo {author} {\bibfnamefont {J.~C.}\ \bibnamefont {Hone}},\ and\
  \bibinfo {author} {\bibfnamefont {A.}~\bibnamefont {Rahimi-Iman}},\
  }\bibfield  {title} {\bibinfo {title} {Direct measurement of the radiative
  pattern of bright and dark excitons and exciton complexes in encapsulated
  tungsten diselenide},\ }\bibfield  {journal} {\bibinfo  {journal} {Scientific
  Reports}\ }\textbf {\bibinfo {volume} {10}},\ \href
  {https://doi.org/10.1038/s41598-020-64838-z} {10.1038/s41598-020-64838-z}
  (\bibinfo {year} {2020})\BibitemShut {NoStop}%
\bibitem [{\citenamefont {Purcell}(1946)}]{Purcell1946}%
  \BibitemOpen
  \bibfield  {author} {\bibinfo {author} {\bibfnamefont {E.~M.}\ \bibnamefont
  {Purcell}},\ }\href@noop {} {\bibinfo {title}
  {{Purcell{\_}1946{\_}SpontaneousEmission.pdf}}} (\bibinfo {year}
  {1946})\BibitemShut {NoStop}%
\bibitem [{\citenamefont {Yablonovitch}(1987)}]{PhysRevLett.58.2059}%
  \BibitemOpen
  \bibfield  {author} {\bibinfo {author} {\bibfnamefont {E.}~\bibnamefont
  {Yablonovitch}},\ }\bibfield  {title} {\bibinfo {title} {{Inhibited
  Spontaneous Emission in Solid-State Physics and Electronics}},\ }\href
  {https://doi.org/10.1103/PhysRevLett.58.2059} {\bibfield  {journal} {\bibinfo
   {journal} {Phys. Rev. Lett.}\ }\textbf {\bibinfo {volume} {58}},\ \bibinfo
  {pages} {2059} (\bibinfo {year} {1987})}\BibitemShut {NoStop}%
\bibitem [{\citenamefont {Khelifa}\ \emph {et~al.}(2020)\citenamefont
  {Khelifa}, \citenamefont {Back}, \citenamefont {Fl\"{o}ry}, \citenamefont
  {Nashashibi}, \citenamefont {Malchow}, \citenamefont {Taniguchi},
  \citenamefont {Watanabe}, \citenamefont {Jain},\ and\ \citenamefont
  {Novotny}}]{Khelifa2020CouplingHeterostructures}%
  \BibitemOpen
  \bibfield  {author} {\bibinfo {author} {\bibfnamefont {R.}~\bibnamefont
  {Khelifa}}, \bibinfo {author} {\bibfnamefont {P.}~\bibnamefont {Back}},
  \bibinfo {author} {\bibfnamefont {N.}~\bibnamefont {Fl\"{o}ry}}, \bibinfo
  {author} {\bibfnamefont {S.}~\bibnamefont {Nashashibi}}, \bibinfo {author}
  {\bibfnamefont {K.}~\bibnamefont {Malchow}}, \bibinfo {author} {\bibfnamefont
  {T.}~\bibnamefont {Taniguchi}}, \bibinfo {author} {\bibfnamefont
  {K.}~\bibnamefont {Watanabe}}, \bibinfo {author} {\bibfnamefont
  {A.}~\bibnamefont {Jain}},\ and\ \bibinfo {author} {\bibfnamefont
  {L.}~\bibnamefont {Novotny}},\ }\bibfield  {title} {\bibinfo {title}
  {Coupling interlayer excitons to whispering gallery modes in van der waals
  heterostructures},\ }\href {https://doi.org/10.1021/acs.nanolett.0c02432}
  {\bibfield  {journal} {\bibinfo  {journal} {Nano Letters}\ }\textbf {\bibinfo
  {volume} {20}},\ \bibinfo {pages} {6155} (\bibinfo {year}
  {2020})}\BibitemShut {NoStop}%
\bibitem [{\citenamefont {Heindel}\ \emph {et~al.}(2017)\citenamefont
  {Heindel}, \citenamefont {Thoma}, \citenamefont {Schwartz}, \citenamefont
  {Schmidgall}, \citenamefont {Gantz}, \citenamefont {Cogan}, \citenamefont
  {Strau{\ss}}, \citenamefont {Schnauber}, \citenamefont {Gschrey},
  \citenamefont {Schulze}, \citenamefont {Strittmatter}, \citenamefont {Rodt},
  \citenamefont {Gershoni},\ and\ \citenamefont
  {Reitzenstein}}]{Heindel2018AccessingMicrolenses}%
  \BibitemOpen
  \bibfield  {author} {\bibinfo {author} {\bibfnamefont {T.}~\bibnamefont
  {Heindel}}, \bibinfo {author} {\bibfnamefont {A.}~\bibnamefont {Thoma}},
  \bibinfo {author} {\bibfnamefont {I.}~\bibnamefont {Schwartz}}, \bibinfo
  {author} {\bibfnamefont {E.~R.}\ \bibnamefont {Schmidgall}}, \bibinfo
  {author} {\bibfnamefont {L.}~\bibnamefont {Gantz}}, \bibinfo {author}
  {\bibfnamefont {D.}~\bibnamefont {Cogan}}, \bibinfo {author} {\bibfnamefont
  {M.}~\bibnamefont {Strau{\ss}}}, \bibinfo {author} {\bibfnamefont
  {P.}~\bibnamefont {Schnauber}}, \bibinfo {author} {\bibfnamefont
  {M.}~\bibnamefont {Gschrey}}, \bibinfo {author} {\bibfnamefont {J.-H.}\
  \bibnamefont {Schulze}}, \bibinfo {author} {\bibfnamefont {A.}~\bibnamefont
  {Strittmatter}}, \bibinfo {author} {\bibfnamefont {S.}~\bibnamefont {Rodt}},
  \bibinfo {author} {\bibfnamefont {D.}~\bibnamefont {Gershoni}},\ and\
  \bibinfo {author} {\bibfnamefont {S.}~\bibnamefont {Reitzenstein}},\
  }\bibfield  {title} {\bibinfo {title} {Accessing the dark exciton spin in
  deterministic quantum-dot microlenses},\ }\href
  {https://doi.org/10.1063/1.5004147} {\bibfield  {journal} {\bibinfo
  {journal} {{APL} Photonics}\ }\textbf {\bibinfo {volume} {2}},\ \bibinfo
  {pages} {121303} (\bibinfo {year} {2017})}\BibitemShut {NoStop}%
\bibitem [{\citenamefont {Andres-Penares}\ \emph {et~al.}(2021)\citenamefont
  {Andres-Penares}, \citenamefont {Habil}, \citenamefont
  {Molina-S{\'{a}}nchez}, \citenamefont {Zapata-Rodr{\'{\i}}guez},
  \citenamefont {Mart{\'{\i}}nez-Pastor},\ and\ \citenamefont
  {S{\'{a}}nchez-Royo}}]{Andres-penaresOut-of-planeMicroresonators}%
  \BibitemOpen
  \bibfield  {author} {\bibinfo {author} {\bibfnamefont {D.}~\bibnamefont
  {Andres-Penares}}, \bibinfo {author} {\bibfnamefont {M.~K.}\ \bibnamefont
  {Habil}}, \bibinfo {author} {\bibfnamefont {A.}~\bibnamefont
  {Molina-S{\'{a}}nchez}}, \bibinfo {author} {\bibfnamefont {C.~J.}\
  \bibnamefont {Zapata-Rodr{\'{\i}}guez}}, \bibinfo {author} {\bibfnamefont
  {J.~P.}\ \bibnamefont {Mart{\'{\i}}nez-Pastor}},\ and\ \bibinfo {author}
  {\bibfnamefont {J.~F.}\ \bibnamefont {S{\'{a}}nchez-Royo}},\ }\bibfield
  {title} {\bibinfo {title} {Out-of-plane trion emission in monolayer {WSe}2
  revealed by whispering gallery modes of dielectric microresonators},\
  }\bibfield  {journal} {\bibinfo  {journal} {Communications Materials}\
  }\textbf {\bibinfo {volume} {2}},\ \href
  {https://doi.org/10.1038/s43246-021-00157-8} {10.1038/s43246-021-00157-8}
  (\bibinfo {year} {2021})\BibitemShut {NoStop}%
\bibitem [{\citenamefont {Liu}\ \emph {et~al.}(2020)\citenamefont {Liu},
  \citenamefont {Chen}, \citenamefont {Tang}, \citenamefont {Vollmer},\ and\
  \citenamefont {Xiao}}]{Liu2021NonlinearFingerprinting}%
  \BibitemOpen
  \bibfield  {author} {\bibinfo {author} {\bibfnamefont {W.}~\bibnamefont
  {Liu}}, \bibinfo {author} {\bibfnamefont {Y.-L.}\ \bibnamefont {Chen}},
  \bibinfo {author} {\bibfnamefont {S.-J.}\ \bibnamefont {Tang}}, \bibinfo
  {author} {\bibfnamefont {F.}~\bibnamefont {Vollmer}},\ and\ \bibinfo {author}
  {\bibfnamefont {Y.-F.}\ \bibnamefont {Xiao}},\ }\bibfield  {title} {\bibinfo
  {title} {Nonlinear sensing with whispering-gallery mode microcavities: From
  label-free detection to spectral fingerprinting},\ }\href
  {https://doi.org/10.1021/acs.nanolett.0c04090} {\bibfield  {journal}
  {\bibinfo  {journal} {Nano Letters}\ }\textbf {\bibinfo {volume} {21}},\
  \bibinfo {pages} {1566} (\bibinfo {year} {2020})}\BibitemShut {NoStop}%
\bibitem [{\citenamefont {Vahala}(2005)}]{Vahala2005}%
  \BibitemOpen
  \bibfield  {author} {\bibinfo {author} {\bibfnamefont {K.~J.}\ \bibnamefont
  {Vahala}},\ }\bibfield  {title} {\bibinfo {title} {{Optical microcavities}},\
  }\href {https://doi.org/10.1109/EQEC.2005.1567517} {\bibfield  {journal}
  {\bibinfo  {journal} {2005 European Quantum Electronics Conference, EQEC
  '05}\ }\textbf {\bibinfo {volume} {2005}},\ \bibinfo {pages} {352} (\bibinfo
  {year} {2005})}\BibitemShut {NoStop}%
\bibitem [{\citenamefont {Ferrera}\ \emph {et~al.}(2008)\citenamefont
  {Ferrera}, \citenamefont {Razzari}, \citenamefont {Duchesne}, \citenamefont
  {Morandotti}, \citenamefont {Yang}, \citenamefont {Liscidini}, \citenamefont
  {Sipe}, \citenamefont {Chu}, \citenamefont {Little},\ and\ \citenamefont
  {Moss}}]{Ferrera2008Low-powerStructures}%
  \BibitemOpen
  \bibfield  {author} {\bibinfo {author} {\bibfnamefont {M.}~\bibnamefont
  {Ferrera}}, \bibinfo {author} {\bibfnamefont {L.}~\bibnamefont {Razzari}},
  \bibinfo {author} {\bibfnamefont {D.}~\bibnamefont {Duchesne}}, \bibinfo
  {author} {\bibfnamefont {R.}~\bibnamefont {Morandotti}}, \bibinfo {author}
  {\bibfnamefont {Z.}~\bibnamefont {Yang}}, \bibinfo {author} {\bibfnamefont
  {M.}~\bibnamefont {Liscidini}}, \bibinfo {author} {\bibfnamefont {J.~E.}\
  \bibnamefont {Sipe}}, \bibinfo {author} {\bibfnamefont {S.}~\bibnamefont
  {Chu}}, \bibinfo {author} {\bibfnamefont {B.~E.}\ \bibnamefont {Little}},\
  and\ \bibinfo {author} {\bibfnamefont {D.~J.}\ \bibnamefont {Moss}},\
  }\bibfield  {title} {\bibinfo {title} {Low-power continuous-wave nonlinear
  optics in doped silica glass integrated waveguide structures},\ }\href
  {https://doi.org/10.1038/nphoton.2008.228} {\bibfield  {journal} {\bibinfo
  {journal} {Nature Photonics}\ }\textbf {\bibinfo {volume} {2}},\ \bibinfo
  {pages} {737} (\bibinfo {year} {2008})}\BibitemShut {NoStop}%
\bibitem [{\citenamefont {Miya}(2000)}]{Miya2000Silica-BasedAnd}%
  \BibitemOpen
  \bibfield  {author} {\bibinfo {author} {\bibfnamefont {T.}~\bibnamefont
  {Miya}},\ }\bibfield  {title} {\bibinfo {title} {Silica-based planar
  lightwave circuits: passive and thermally active devices},\ }\href
  {https://doi.org/10.1109/2944.826871} {\bibfield  {journal} {\bibinfo
  {journal} {{IEEE} Journal of Selected Topics in Quantum Electronics}\
  }\textbf {\bibinfo {volume} {6}},\ \bibinfo {pages} {38} (\bibinfo {year}
  {2000})}\BibitemShut {NoStop}%
\bibitem [{\citenamefont {Ying}\ \emph {et~al.}(2018)\citenamefont {Ying},
  \citenamefont {Wang}, \citenamefont {Zhao}, \citenamefont {Dhar},
  \citenamefont {Pan}, \citenamefont {Soref},\ and\ \citenamefont
  {Chen}}]{Ying2018ComparisonPhotonics}%
  \BibitemOpen
  \bibfield  {author} {\bibinfo {author} {\bibfnamefont {Z.}~\bibnamefont
  {Ying}}, \bibinfo {author} {\bibfnamefont {Z.}~\bibnamefont {Wang}}, \bibinfo
  {author} {\bibfnamefont {Z.}~\bibnamefont {Zhao}}, \bibinfo {author}
  {\bibfnamefont {S.}~\bibnamefont {Dhar}}, \bibinfo {author} {\bibfnamefont
  {D.~Z.}\ \bibnamefont {Pan}}, \bibinfo {author} {\bibfnamefont
  {R.}~\bibnamefont {Soref}},\ and\ \bibinfo {author} {\bibfnamefont {R.~T.}\
  \bibnamefont {Chen}},\ }\bibfield  {title} {\bibinfo {title} {Comparison of
  microrings and microdisks for high-speed optical modulation in silicon
  photonics},\ }\href {https://doi.org/10.1063/1.5019590} {\bibfield  {journal}
  {\bibinfo  {journal} {Applied Physics Letters}\ }\textbf {\bibinfo {volume}
  {112}},\ \bibinfo {pages} {111108} (\bibinfo {year} {2018})}\BibitemShut
  {NoStop}%
\bibitem [{\citenamefont {Li}\ \emph {et~al.}(2015)\citenamefont {Li},
  \citenamefont {Liu}, \citenamefont {Jiang}, \citenamefont {Yang},
  \citenamefont {Ma}, \citenamefont {Wu},\ and\ \citenamefont
  {Xiao}}]{Li2015High-Chip}%
  \BibitemOpen
  \bibfield  {author} {\bibinfo {author} {\bibfnamefont {G.}~\bibnamefont
  {Li}}, \bibinfo {author} {\bibfnamefont {P.}~\bibnamefont {Liu}}, \bibinfo
  {author} {\bibfnamefont {X.}~\bibnamefont {Jiang}}, \bibinfo {author}
  {\bibfnamefont {C.}~\bibnamefont {Yang}}, \bibinfo {author} {\bibfnamefont
  {J.}~\bibnamefont {Ma}}, \bibinfo {author} {\bibfnamefont {H.}~\bibnamefont
  {Wu}},\ and\ \bibinfo {author} {\bibfnamefont {M.}~\bibnamefont {Xiao}},\
  }\bibfield  {title} {\bibinfo {title} {High-q silica microdisk optical
  resonators with large wedge angles on a silicon chip},\ }\href
  {https://doi.org/10.1364/prj.3.000279} {\bibfield  {journal} {\bibinfo
  {journal} {Photonics Research}\ }\textbf {\bibinfo {volume} {3}},\ \bibinfo
  {pages} {279} (\bibinfo {year} {2015})}\BibitemShut {NoStop}%
\bibitem [{\citenamefont {Chen}\ \emph {et~al.}(2018)\citenamefont {Chen},
  \citenamefont {Wan}, \citenamefont {Chandrahalim}, \citenamefont {Zhou},
  \citenamefont {Zhang}, \citenamefont {Cho}, \citenamefont {Mei},
  \citenamefont {Yoshioka}, \citenamefont {Tian}, \citenamefont {Nishimura},
  \citenamefont {Fan}, \citenamefont {Guo},\ and\ \citenamefont
  {Oki}}]{Chen2018EffectsLasers}%
  \BibitemOpen
  \bibfield  {author} {\bibinfo {author} {\bibfnamefont {C.}~\bibnamefont
  {Chen}}, \bibinfo {author} {\bibfnamefont {L.}~\bibnamefont {Wan}}, \bibinfo
  {author} {\bibfnamefont {H.}~\bibnamefont {Chandrahalim}}, \bibinfo {author}
  {\bibfnamefont {J.}~\bibnamefont {Zhou}}, \bibinfo {author} {\bibfnamefont
  {H.}~\bibnamefont {Zhang}}, \bibinfo {author} {\bibfnamefont
  {S.}~\bibnamefont {Cho}}, \bibinfo {author} {\bibfnamefont {T.}~\bibnamefont
  {Mei}}, \bibinfo {author} {\bibfnamefont {H.}~\bibnamefont {Yoshioka}},
  \bibinfo {author} {\bibfnamefont {H.}~\bibnamefont {Tian}}, \bibinfo {author}
  {\bibfnamefont {N.}~\bibnamefont {Nishimura}}, \bibinfo {author}
  {\bibfnamefont {X.}~\bibnamefont {Fan}}, \bibinfo {author} {\bibfnamefont
  {L.~J.}\ \bibnamefont {Guo}},\ and\ \bibinfo {author} {\bibfnamefont
  {Y.}~\bibnamefont {Oki}},\ }\bibfield  {title} {\bibinfo {title} {Effects of
  edge inclination angles on whispering-gallery modes in printable wedge
  microdisk lasers},\ }\href {https://doi.org/10.1364/oe.26.000233} {\bibfield
  {journal} {\bibinfo  {journal} {Optics Express}\ }\textbf {\bibinfo {volume}
  {26}},\ \bibinfo {pages} {233} (\bibinfo {year} {2018})}\BibitemShut
  {NoStop}%
\bibitem [{\citenamefont {Ma}\ \emph {et~al.}(2019)\citenamefont {Ma},
  \citenamefont {Xiao}, \citenamefont {Gu}, \citenamefont {Li}, \citenamefont
  {Cheng}, \citenamefont {He}, \citenamefont {Jiang},\ and\ \citenamefont
  {Xiao}}]{Iyang2019VisibleAngle}%
  \BibitemOpen
  \bibfield  {author} {\bibinfo {author} {\bibfnamefont {J.}~\bibnamefont
  {Ma}}, \bibinfo {author} {\bibfnamefont {L.}~\bibnamefont {Xiao}}, \bibinfo
  {author} {\bibfnamefont {J.}~\bibnamefont {Gu}}, \bibinfo {author}
  {\bibfnamefont {H.}~\bibnamefont {Li}}, \bibinfo {author} {\bibfnamefont
  {X.}~\bibnamefont {Cheng}}, \bibinfo {author} {\bibfnamefont
  {G.}~\bibnamefont {He}}, \bibinfo {author} {\bibfnamefont {X.}~\bibnamefont
  {Jiang}},\ and\ \bibinfo {author} {\bibfnamefont {M.}~\bibnamefont {Xiao}},\
  }\bibfield  {title} {\bibinfo {title} {Visible kerr comb generation in a
  high-q silica microdisk resonator with a large wedge angle},\ }\href
  {https://doi.org/10.1364/prj.7.000573} {\bibfield  {journal} {\bibinfo
  {journal} {Photonics Research}\ }\textbf {\bibinfo {volume} {7}},\ \bibinfo
  {pages} {573} (\bibinfo {year} {2019})}\BibitemShut {NoStop}%
\bibitem [{\citenamefont {Novotny}\ and\ \citenamefont
  {Hecht}(2006)}]{Novotny2006}%
  \BibitemOpen
  \bibfield  {author} {\bibinfo {author} {\bibfnamefont {L.}~\bibnamefont
  {Novotny}}\ and\ \bibinfo {author} {\bibfnamefont {B.}~\bibnamefont
  {Hecht}},\ }\href {https://doi.org/https://doi.org/10.1017/CBO9780511813535}
  {\emph {\bibinfo {title} {{Principles of Nano-Optics}}}}\ (\bibinfo
  {publisher} {Cambridge University Press},\ \bibinfo {year}
  {2006})\BibitemShut {NoStop}%
\bibitem [{\citenamefont {Park}\ \emph {et~al.}(2016)\citenamefont {Park},
  \citenamefont {Kim},\ and\ \citenamefont {Jeong}}]{Park2016}%
  \BibitemOpen
  \bibfield  {author} {\bibinfo {author} {\bibfnamefont {K.-W.}\ \bibnamefont
  {Park}}, \bibinfo {author} {\bibfnamefont {J.}~\bibnamefont {Kim}},\ and\
  \bibinfo {author} {\bibfnamefont {K.}~\bibnamefont {Jeong}},\ }\bibfield
  {title} {\bibinfo {title} {Non-hermitian hamiltonian and lamb shift in
  circular dielectric microcavity},\ }\href
  {https://doi.org/10.1016/j.optcom.2016.02.001} {\bibfield  {journal}
  {\bibinfo  {journal} {Optics Communications}\ }\textbf {\bibinfo {volume}
  {368}},\ \bibinfo {pages} {190} (\bibinfo {year} {2016})}\BibitemShut
  {NoStop}%
\bibitem [{\citenamefont {Barnes}\ \emph {et~al.}(2020)\citenamefont {Barnes},
  \citenamefont {Horsley},\ and\ \citenamefont {Vos}}]{Barnes2020}%
  \BibitemOpen
  \bibfield  {author} {\bibinfo {author} {\bibfnamefont {W.~L.}\ \bibnamefont
  {Barnes}}, \bibinfo {author} {\bibfnamefont {S.~A.~R.}\ \bibnamefont
  {Horsley}},\ and\ \bibinfo {author} {\bibfnamefont {W.~L.}\ \bibnamefont
  {Vos}},\ }\bibfield  {title} {\bibinfo {title} {Classical antennae, quantum
  emitters, and densities of optical states},\ }\bibfield  {journal} {\bibinfo
  {journal} {Journal of Optics}\ }\href
  {https://doi.org/10.1088/2040-8986/ab7b01} {10.1088/2040-8986/ab7b01}
  (\bibinfo {year} {2020})\BibitemShut {NoStop}%
\bibitem [{\citenamefont {Malic}\ \emph {et~al.}(2018)\citenamefont {Malic},
  \citenamefont {Selig}, \citenamefont {Feierabend}, \citenamefont {Brem},
  \citenamefont {Christiansen}, \citenamefont {Wendler}, \citenamefont
  {Knorr},\ and\ \citenamefont {Bergh\"auser}}]{PhysRevMaterials.2.014002}%
  \BibitemOpen
  \bibfield  {author} {\bibinfo {author} {\bibfnamefont {E.}~\bibnamefont
  {Malic}}, \bibinfo {author} {\bibfnamefont {M.}~\bibnamefont {Selig}},
  \bibinfo {author} {\bibfnamefont {M.}~\bibnamefont {Feierabend}}, \bibinfo
  {author} {\bibfnamefont {S.}~\bibnamefont {Brem}}, \bibinfo {author}
  {\bibfnamefont {D.}~\bibnamefont {Christiansen}}, \bibinfo {author}
  {\bibfnamefont {F.}~\bibnamefont {Wendler}}, \bibinfo {author} {\bibfnamefont
  {A.}~\bibnamefont {Knorr}},\ and\ \bibinfo {author} {\bibfnamefont
  {G.}~\bibnamefont {Bergh\"auser}},\ }\bibfield  {title} {\bibinfo {title}
  {Dark excitons in transition metal dichalcogenides},\ }\href
  {https://doi.org/10.1103/PhysRevMaterials.2.014002} {\bibfield  {journal}
  {\bibinfo  {journal} {Phys. Rev. Materials}\ }\textbf {\bibinfo {volume}
  {2}},\ \bibinfo {pages} {014002} (\bibinfo {year} {2018})}\BibitemShut
  {NoStop}%
\bibitem [{\citenamefont {Schwartz}\ \emph {et~al.}(2015)\citenamefont
  {Schwartz}, \citenamefont {Schmidgall}, \citenamefont {Gantz}, \citenamefont
  {Cogan}, \citenamefont {Bordo}, \citenamefont {Don}, \citenamefont
  {Zielinski},\ and\ \citenamefont {Gershoni}}]{PhysRevX.5.011009}%
  \BibitemOpen
  \bibfield  {author} {\bibinfo {author} {\bibfnamefont {I.}~\bibnamefont
  {Schwartz}}, \bibinfo {author} {\bibfnamefont {E.~R.}\ \bibnamefont
  {Schmidgall}}, \bibinfo {author} {\bibfnamefont {L.}~\bibnamefont {Gantz}},
  \bibinfo {author} {\bibfnamefont {D.}~\bibnamefont {Cogan}}, \bibinfo
  {author} {\bibfnamefont {E.}~\bibnamefont {Bordo}}, \bibinfo {author}
  {\bibfnamefont {Y.}~\bibnamefont {Don}}, \bibinfo {author} {\bibfnamefont
  {M.}~\bibnamefont {Zielinski}},\ and\ \bibinfo {author} {\bibfnamefont
  {D.}~\bibnamefont {Gershoni}},\ }\bibfield  {title} {\bibinfo {title}
  {Deterministic writing and control of the dark exciton spin using single
  short optical pulses},\ }\href {https://doi.org/10.1103/PhysRevX.5.011009}
  {\bibfield  {journal} {\bibinfo  {journal} {Phys. Rev. X}\ }\textbf {\bibinfo
  {volume} {5}},\ \bibinfo {pages} {011009} (\bibinfo {year}
  {2015})}\BibitemShut {NoStop}%
\bibitem [{\citenamefont {Jha}\ \emph {et~al.}(2018)\citenamefont {Jha},
  \citenamefont {Shitrit}, \citenamefont {Ren}, \citenamefont {Wang},\ and\
  \citenamefont {Zhang}}]{PhysRevLett.121.116102}%
  \BibitemOpen
  \bibfield  {author} {\bibinfo {author} {\bibfnamefont {P.~K.}\ \bibnamefont
  {Jha}}, \bibinfo {author} {\bibfnamefont {N.}~\bibnamefont {Shitrit}},
  \bibinfo {author} {\bibfnamefont {X.}~\bibnamefont {Ren}}, \bibinfo {author}
  {\bibfnamefont {Y.}~\bibnamefont {Wang}},\ and\ \bibinfo {author}
  {\bibfnamefont {X.}~\bibnamefont {Zhang}},\ }\bibfield  {title} {\bibinfo
  {title} {Spontaneous exciton valley coherence in transition metal
  dichalcogenide monolayers interfaced with an anisotropic metasurface},\
  }\href {https://doi.org/10.1103/PhysRevLett.121.116102} {\bibfield  {journal}
  {\bibinfo  {journal} {Phys. Rev. Lett.}\ }\textbf {\bibinfo {volume} {121}},\
  \bibinfo {pages} {116102} (\bibinfo {year} {2018})}\BibitemShut {NoStop}%
\bibitem [{\citenamefont {Nalabothula}\ \emph {et~al.}(2020)\citenamefont
  {Nalabothula}, \citenamefont {Jha}, \citenamefont {Low},\ and\ \citenamefont
  {Kumar}}]{PhysRevB.102.045416}%
  \BibitemOpen
  \bibfield  {author} {\bibinfo {author} {\bibfnamefont {M.}~\bibnamefont
  {Nalabothula}}, \bibinfo {author} {\bibfnamefont {P.~K.}\ \bibnamefont
  {Jha}}, \bibinfo {author} {\bibfnamefont {T.}~\bibnamefont {Low}},\ and\
  \bibinfo {author} {\bibfnamefont {A.}~\bibnamefont {Kumar}},\ }\bibfield
  {title} {\bibinfo {title} {Engineering valley quantum interference in
  anisotropic van der waals heterostructures},\ }\href
  {https://doi.org/10.1103/PhysRevB.102.045416} {\bibfield  {journal} {\bibinfo
   {journal} {Phys. Rev. B}\ }\textbf {\bibinfo {volume} {102}},\ \bibinfo
  {pages} {045416} (\bibinfo {year} {2020})}\BibitemShut {NoStop}%
\bibitem [{\citenamefont {Schneider}\ \emph {et~al.}(2018)\citenamefont
  {Schneider}, \citenamefont {Glazov}, \citenamefont {Korn}, \citenamefont
  {H\"{o}fling},\ and\ \citenamefont {Urbaszek}}]{Schneider2018}%
  \BibitemOpen
  \bibfield  {author} {\bibinfo {author} {\bibfnamefont {C.}~\bibnamefont
  {Schneider}}, \bibinfo {author} {\bibfnamefont {M.~M.}\ \bibnamefont
  {Glazov}}, \bibinfo {author} {\bibfnamefont {T.}~\bibnamefont {Korn}},
  \bibinfo {author} {\bibfnamefont {S.}~\bibnamefont {H\"{o}fling}},\ and\
  \bibinfo {author} {\bibfnamefont {B.}~\bibnamefont {Urbaszek}},\ }\bibfield
  {title} {\bibinfo {title} {Two-dimensional semiconductors in the regime of
  strong light-matter coupling},\ }\bibfield  {journal} {\bibinfo  {journal}
  {Nature Communications}\ }\textbf {\bibinfo {volume} {9}},\ \href
  {https://doi.org/10.1038/s41467-018-04866-6} {10.1038/s41467-018-04866-6}
  (\bibinfo {year} {2018})\BibitemShut {NoStop}%
\bibitem [{\citenamefont {Koshelev}\ \emph {et~al.}(2018)\citenamefont
  {Koshelev}, \citenamefont {Sychev}, \citenamefont {Sadrieva}, \citenamefont
  {Bogdanov},\ and\ \citenamefont {Iorsh}}]{PhysRevB.98.161113}%
  \BibitemOpen
  \bibfield  {author} {\bibinfo {author} {\bibfnamefont {K.~L.}\ \bibnamefont
  {Koshelev}}, \bibinfo {author} {\bibfnamefont {S.~K.}\ \bibnamefont
  {Sychev}}, \bibinfo {author} {\bibfnamefont {Z.~F.}\ \bibnamefont
  {Sadrieva}}, \bibinfo {author} {\bibfnamefont {A.~A.}\ \bibnamefont
  {Bogdanov}},\ and\ \bibinfo {author} {\bibfnamefont {I.~V.}\ \bibnamefont
  {Iorsh}},\ }\bibfield  {title} {\bibinfo {title} {Strong coupling between
  excitons in transition metal dichalcogenides and optical bound states in the
  continuum},\ }\href {https://doi.org/10.1103/PhysRevB.98.161113} {\bibfield
  {journal} {\bibinfo  {journal} {Phys. Rev. B}\ }\textbf {\bibinfo {volume}
  {98}},\ \bibinfo {pages} {161113} (\bibinfo {year} {2018})}\BibitemShut
  {NoStop}%
\bibitem [{\citenamefont {Jiang}\ \emph
  {et~al.}(2021{\natexlab{b}})\citenamefont {Jiang}, \citenamefont {Chen},
  \citenamefont {Zheng}, \citenamefont {Zheng},\ and\ \citenamefont
  {Pan}}]{Jiang2021}%
  \BibitemOpen
  \bibfield  {author} {\bibinfo {author} {\bibfnamefont {Y.}~\bibnamefont
  {Jiang}}, \bibinfo {author} {\bibfnamefont {S.}~\bibnamefont {Chen}},
  \bibinfo {author} {\bibfnamefont {W.}~\bibnamefont {Zheng}}, \bibinfo
  {author} {\bibfnamefont {B.}~\bibnamefont {Zheng}},\ and\ \bibinfo {author}
  {\bibfnamefont {A.}~\bibnamefont {Pan}},\ }\bibfield  {title} {\bibinfo
  {title} {Interlayer exciton formation, relaxation, and transport in {TMD} van
  der waals heterostructures},\ }\bibfield  {journal} {\bibinfo  {journal}
  {Light: Science {\&} Applications}\ }\textbf {\bibinfo {volume} {10}},\ \href
  {https://doi.org/10.1038/s41377-021-00500-1} {10.1038/s41377-021-00500-1}
  (\bibinfo {year} {2021}{\natexlab{b}})\BibitemShut {NoStop}%
\bibitem [{\citenamefont {Sohoni}\ \emph {et~al.}(2020)\citenamefont {Sohoni},
  \citenamefont {Jha}, \citenamefont {Nalabothula},\ and\ \citenamefont
  {Kumar}}]{Sohoni2020}%
  \BibitemOpen
  \bibfield  {author} {\bibinfo {author} {\bibfnamefont {M.}~\bibnamefont
  {Sohoni}}, \bibinfo {author} {\bibfnamefont {P.~K.}\ \bibnamefont {Jha}},
  \bibinfo {author} {\bibfnamefont {M.}~\bibnamefont {Nalabothula}},\ and\
  \bibinfo {author} {\bibfnamefont {A.}~\bibnamefont {Kumar}},\ }\bibfield
  {title} {\bibinfo {title} {Interlayer exciton valleytronics in bilayer
  heterostructures interfaced with a phase gradient metasurface},\ }\href
  {https://doi.org/10.1063/5.0015087} {\bibfield  {journal} {\bibinfo
  {journal} {Applied Physics Letters}\ }\textbf {\bibinfo {volume} {117}},\
  \bibinfo {pages} {121101} (\bibinfo {year} {2020})}\BibitemShut {NoStop}%
\end{thebibliography}%

\end{document}